\pdfoutput=1
\documentclass[aps,prl,reprint,amsmath,amssymb,superscriptaddress]{revtex4-1}
\usepackage{bm}
\usepackage{graphicx}
\usepackage[colorlinks]{hyperref}
\usepackage{mathrsfs}
\usepackage{times}
\usepackage[dvipsnames]{xcolor}
\usepackage{float}
\usepackage{braket}

\newcommand{\bit}{\begin{enumerate}}
	\newcommand{\eit}{\end{enumerate}}

\def\em{\it}
\definecolor{bananayellow}{rgb}{1.0, 0.88, 0.21}
\definecolor{straw}{rgb}{0.32, 0.28, 0.1}

\begin{document}
	\title{Suppression of inter-band heating for random  driving }

		\author{Hongzheng Zhao}
\affiliation{\small Max-Planck-Institut f{\"u}r Physik komplexer Systeme, N{\"o}thnitzer Stra{\ss}e 38, 01187 Dresden, Germany}
	\affiliation{\small Blackett Laboratory, Imperial College London, London SW7 2AZ, United Kingdom}

		\author{Johannes Knolle  }
	\affiliation{Department of Physics TQM, Technische Universit{\"a}t M{\"u}nchen, James-Franck-Stra{\ss}e 1, D-85748 Garching, Germany}
	\affiliation{Munich Center for Quantum Science and Technology (MCQST), 80799 Munich, Germany}
	\affiliation{\small Blackett Laboratory, Imperial College London, London SW7 2AZ, United Kingdom}
	
	\author{Roderich Moessner}
		\affiliation{\small Max-Planck-Institut f{\"u}r Physik komplexer Systeme, N{\"o}thnitzer Stra{\ss}e 38, 01187 Dresden, Germany}
	
	\author{Florian Mintert}
	\affiliation{\small Blackett Laboratory, Imperial College London, London SW7 2AZ, United Kingdom}
	
	\begin{abstract}
Heating to high-lying states strongly limits the experimental observation of driving induced non-equilibrium phenomena, particularly when the drive has a broad spectrum. Here we show that, for entire families of structured random drives known as random multipolar drives, particle excitation to higher bands can be well controlled even away from a  high-frequency driving regime. This opens a window for observing drive-induced phenomena in a long-lived prethermal regime in the lowest band.
\end{abstract}

	\maketitle
{\it Introduction.---}
Quantum simulators hold the promise of exploring physics that is far beyond the capabilities of any conceivable classical simulation.
A crucial tool is driving quantum systems, since this can result in drastic modifications of their properties~\cite{moessner2017equilibration,wintersperger2020realization,yang2020observation,martinez2016real,arute2019quantum,song2021realizing}.
In particular, periodic driving is used to {\em Floquet engineer} either time-independent effective models with exotic physical processes~\cite{oka2019floquet,bukov2015universal} or time-dependent models exhibiting non-equilibrium phases of matter which do not exist in static systems~\cite{eckardt2017colloquium}.
Prominent examples include Floquet discrete time crystals~\cite{khemani2016phase,else2016floquet}, artificial gauge fields for neutral particles~\cite{struck2013engineering,aidelsburger2013realization,gross2017quantum} and novel topological phases of matter~\cite{nathan2019anomalous,nathan2017quantized,kitagawa2010topological}.

Actual limitations of successful Floquet engineering result from heating~\cite{rubio2020floquet,reitter2017interaction,viebahn2021suppressing}:
in addition to the desired modification of quantum dynamics, driving also induces undesired transitions to excited states.
In the case of lattice models with a well-defined band structure, the heating effects can be categorized into two classes.
Excitation to higher bands (inter-band heating) inevitably leads to particle loss from the lowest band that is typically used for the quantum simulation~\cite{viebahn2021suppressing};
and even if inter-band heating is sufficiently slow, intra-band heating lets any generic many-body system thermalize to a high-temperature state within the lowest band~\cite{lazarides2014equilibrium,abanin2015exponentially,mori2018thermalization}.  

Quantum simulations thus typically rely on the existence of a sweet spot of drives that can induce the desired dynamics without causing too much heating.
Periodic driving allows to suppress intra-band heating in terms of a sufficiently high fundamental frequency~\cite{kuwahara2016floquet,abanin2017rigorous,eckardt2015high,machado2020long,pizzi2021classical}.
Many-body resonances in the low-energy subspace are exponentially suppressed, so that a quasi-stationary prethermal regime can exist.
Within the lowest-band approximation, the lifetime of the prethermal regime can be arbitrarily long  for a sufficiently fast drive.
In realistic many-band systems, however, the transitions to higher bands that can become resonant with high-frequency driving pose severe limitations to the practically accessible driving spectra~\cite{sun2020optimal}.
Due to the discrete spectrum of periodic driving, this is not an unsurmountable obstacle, and there are driving frequencies that give access to long-lived Floquet prethermalization in a series of quantum systems, including NV centers~\cite{beatrez2021floquet}, trapped ions~\cite{kyprianidis2021observation}, NMR~\cite{peng2021floquet} and Hubbard type systems~\cite{rubio2020floquet,viebahn2021suppressing}.

Due to the broader spectra of aperiodic drives, it is much more challenging to find driving patterns that realize desired dynamics without excessive heating~\cite{martin2017topological,long2021many,friedman2020topological,zhao2021random,nandy2017aperiodically,crowley2019topological,nathan2020quasiperiodic,malz2021topological,dumitrescu2018logarithmically,zhao2019floquet}.
Prethermalization can occur also for quasi-periodic~\cite{else2020long,mori2021rigorous} and structured random drives~\cite{zhao2021random},
but inter-band heating with aperiodic driving is a largely uncharted territory.
The complicated spectrum can potentially induce stronger particle loss than with periodic drives, and the existence of a long-lived prethermal regime is not garantueed.

In this work, we show that random multipolar driving (RMD)~\cite{zhao2021random,mori2021rigorous}, which interpolates between random and quasi-periodic drives, can still provide access to long-lived prethermal phenomena. 
We focus on the experimentally relevant Bose-Hubbard model (BHM) and discover a highly non-monotonic dependence of the inter-band heating on the driving frequency.
Most importantly, within specific frequency windows, inter-band heating can be exceptionally suppressed, when a single particle can approximately complete Rabi-like cycles between the bands and recover its initial state at stroboscopic times. 
Crucially, such suppression extends to situations where many-body interactions are present as long as the band gap is sufficiently large. 

In the following, we first introduce the model as well as the RMD protocol. We confirm long-lived prethermalization within the lowest band in the absence of particle loss to the higher band. We then investigate the non-monotonic inter-band heating profile which can be explained via a simple tractable theory. We finally show that the observation of the prethermal phenomenon where particle loss is well controlled is within experimental reach of state-of-the-art quantum simulators
in suitable frequency windows.
 
{\it The model.---}  The dynamics of spinless bosons within a single band is characterized by the Hamiltonian
\begin{eqnarray}
\begin{aligned}
\hat{H}(J,U,h_l,\vec b)=-J&\sum_{l}\left(\hat{b}_{l+1}^{\dagger} \hat{b}_{l}+\text {h.c.}\right)\\
&+\frac{U}{2} \sum_{l} \hat{n}_{l}\left(\hat{n}_{l}-1\right) +\sum_{l}h_l\hat{n}_{l}, 
\end{aligned}
\end{eqnarray}
with hopping rate $J$, interaction constant $U$, onsite energies $h_l$ and bosonic creation operators $\hat{b}_{l}^{\dagger}$ and number operators $\hat n_l=\hat{b}_{l}^{\dagger}\hat{b}_{l}$ on sites labelled by $l$ in the corresponding band.

The low-energy subspace is the s-band with Hamiltonian $\hat H_{\mathrm{s}}=\hat{H}(J_s,U_s,h_l,\vec b_s)$.
Since the hopping rate can be directly tuned via the lattice depth, it is common to realize a time-dependent Hamiltonian in terms of a time-dependent hopping rate.
In the following, we thus consider the piecewise constant drive of the s-band BHM $\hat H_{\mathrm{s},\pm}=\hat{H}(J_s\pm\delta J_s,U_s, h_l,\vec b_s)$   for the $n-$RMD protocol specified later.

Modulation of the lattice depth excites particles to higher bands. As the first excited (p) band has odd parity, transitions from the even-parity s-band
to higher bands are dominated by the coupling to the second excited (d) band~\cite{rubio2020floquet}.
Including the d-band in terms of the Hamiltonian $\hat H_{\mathrm{d}}=\hat{H}(J_d,U_d,h_l+\Delta,\vec b_d)$ with an energy gap $\Delta$ to the s-band is the central step in going beyond the lowest-band-approximation.
This energy gap  is normally much larger than the s-band hopping rate $J_s$; in a deep lattice it can even be larger by two orders of magnitude~\cite{rubio2020floquet,viebahn2021suppressing,sun2020optimal}. 
The Hamiltonian
\begin{eqnarray}
\label{eq.inter_band}
\hat{H}_{\mathrm{sd}}=U_{sd} \sum_{l}\left[2 \hat{n}_{s l} \hat{n}_{d l}+\frac{1}{2}\left(\hat{b}_{d l}^{\dagger} \hat{b}_{d l}^{\dagger} \hat{b}_{s l} \hat{b}_{s l}+\mathrm{h.c.}\right)\right]\ ,
\end{eqnarray}
where the index $sl$ or $dl$ labels site $l$ in the s or d-band, respectively, captures the inter-band interaction including an on-site density-density interaction and simultaneous hopping of two particles  between the bands.
Modulation of the lattice depth changes not only the hopping rate in the s-band, but it also induces single particle inter-band transitions to the d-band ~\cite{sun2020optimal} as described by the Hamiltonian
\begin{eqnarray}
\label{eq.s-pdrive}
\hat{H}_{\mathrm{tr}}=\eta\delta J_s \sum_{l}\hat{b}_{d l}^{\dagger} \hat{b}_{s l}+\mathrm{h.c.}, 
\end{eqnarray} 
with a dimensionless transition ratio $\eta$. 
Strictly speaking the lattice modulation will also let parameters
like the hopping rate in the d-band or the interaction amplitudes vary in time.
However,
this time-dependency will not lead to sizeable effects as long as the band gap is sufficiently large.
Therefore, in addition to the s-band drive, we only consider the driving dependent single particle transitions. 

Our goal is to investigate the heating process of the two-band BHM subject to the RMD protocol~\cite{zhao2021random}, which can be defined as a sequence of the two piecewise constant Hamiltonians  $\hat{H}_{\pm}=\hat{H}_{\mathrm{s},\pm}+\hat{H}_\mathrm{d}+\hat{H}_\mathrm{sd}\pm \hat{H}_\mathrm{tr}.$
A period of deterministic dynamics over a time-window of length $2^{n}T$ can be defined recursively via the time evolution operators $U_{n}^{\pm}=U_{n-1}^{\mp} U_{n-1}^{\pm}$,
with $U_0^{\pm}=e^{-i T \hat H_\pm}$.
Temporal randomness can be introduced by concatenating dynamics over several such time-windows with the operator $U_{n}^{\pm}$  chosen at random.

Long-lived prethermalization governed by the effective Hamiltonian $\hat{H}_{\mathrm{eff}}=\hat{H}_++\hat{H}_-$ is predicted~\cite{zhao2021random,mori2021rigorous} in the rapid driving regime where the characteristic driving frequency $1/T$ is the dominant energy scale.
The scaling $\tau_{\mathrm{pre}}\sim T^{-(2n+1)}$ of the prethermal lifetime 
has been observed numerically in non-integrable spin models~\cite{zhao2021random,mori2021rigorous}.
In the following, we first establish similar prethermal phenomena in the lowest s-band. Taking into account multiple-bands, we then confirm the characteristic scaling of the prethermal lifetime even though $1/T$ is only larger than the energy scales of the lower band but smaller than the gap to higher ones.

\begin{figure}
	\centering
	\includegraphics[width=0.99\linewidth]{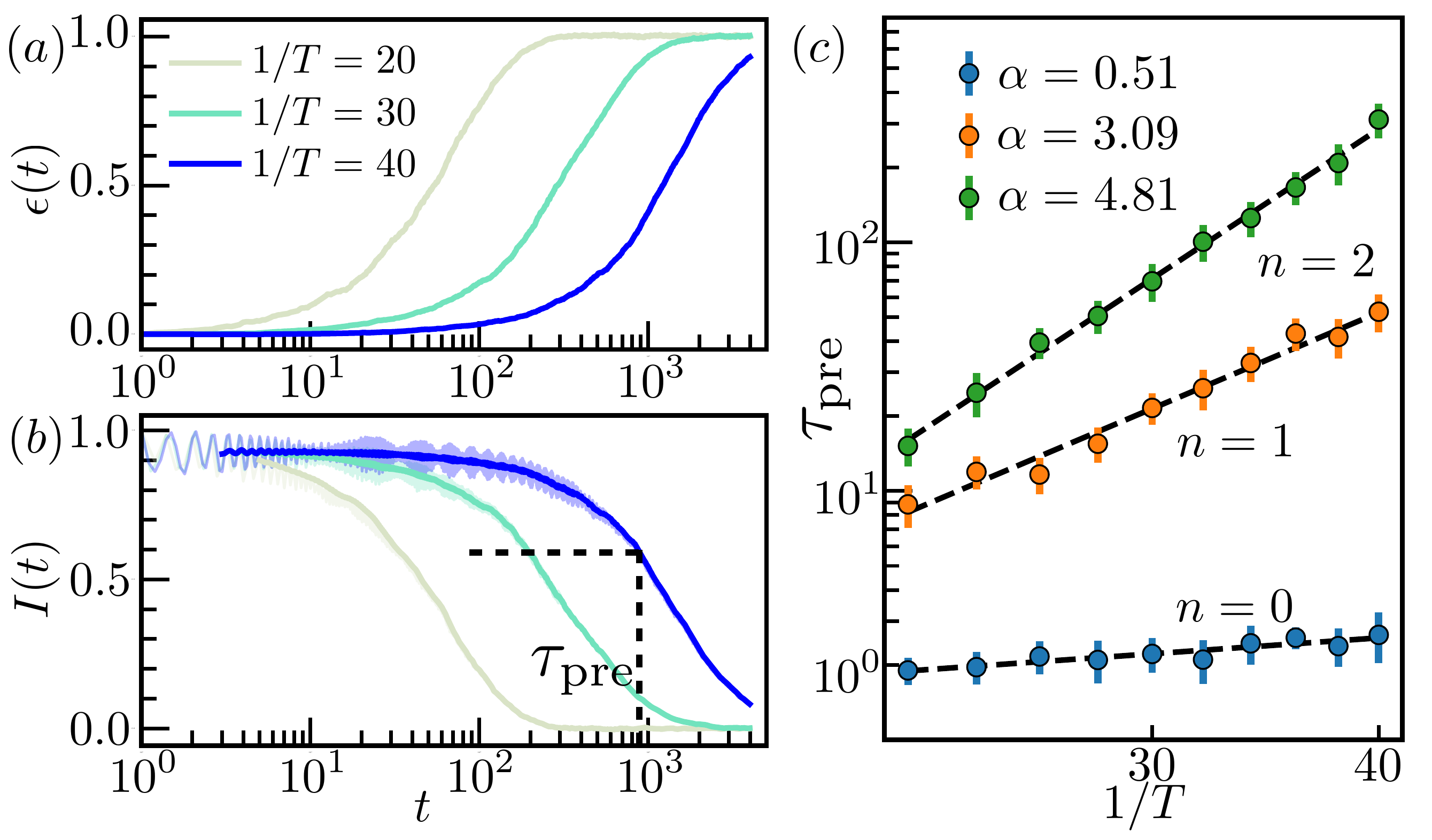}
	\caption{Dynamics of the energy (a) and imbalance (b) for the s-band BHM. A prethermal plateau appears in both panels with a lifetime $\tau_{\mathrm{pre}}$ increasing with driving frequency $1/T$. (c) Algebraic scaling {$\tau_{\mathrm{pre}}\sim T^{-(2n+1)}$} can be observed for $n\geq 1$, whereas for $n=0$, heating rapidly happens ($t\sim 1$). The follow parameters are used for numerical simulation $J_s=1,\delta J_s=0.9,U_s=5,\mu=10.$}
	\label{fig:dynamics_clean}
\end{figure}

{\it Prethermalization in the s-band.---}
The existence of a prethermal regime within the lowest-band approximation is indicated in Fig.~\ref{fig:dynamics_clean}
with dynamics following an average over 100 instances of $2-$RMD driving with periodic boundary conditions. The initial state is chosen as a density-wave state in the lattice of $L=12$ sites and total particle number $N_{\mathrm{tot}}=6$.

Inset (a) depicts the normalized energy 
$\varepsilon\left(t\right) \equiv (E_t-E_{0})/(E_{\infty}-E_{0})$,
with the energy expectation $E_t=\bra{\Psi_t}\hat{H}_{\mathrm{eff}}\ket{\Psi_t}$ taken with respect to the effective Hamiltonian,
the infinite temperature energy $E_{\infty}=\operatorname{Tr}\hat{H}_{\mathrm{eff}}/D$ where $D$ is the Hilbert space dimension
and the energy  $E_{0}=\bra{\Psi_0}\hat{H}_{\mathrm{eff}}\ket{\Psi_0}$ of the initial state $\ket{\Psi_0}$.
Inset (b) shows the imbalance between the occupation on even and odd sites labeled by $sl$ in the s-band
$
{I}(t) =\frac{2}{L} (\sum_{l,\mathrm{even}} \langle\hat{n}_{sl}\rangle_t-\sum_{l,\mathrm{odd}} \langle \hat{n}_{sl}\rangle_t)\ ,
$ with the occupation number expectation value $\langle \hat{n}_{sl}\rangle_t=\bra{\Psi_t}\hat{n}_{sl}\ket{\Psi_t}$.

For the three representative values of the driving frequency $1/T$ shown in Fig.~\ref{fig:dynamics_clean} (a) and (b), the values of $\varepsilon$ and $I$ remain approximately constant during the prethermal plateau. Eventually they approach their infinite temperature value on some finite time scale, which depends on $1/T$, with fast driving (large values of $1/T$) favoring slow thermalization.
The transition between the initial and final values of $\varepsilon$ at any given value of $1/T$ takes place in the same time-window as the corresponding transition of ${I}$, indicating that the definition of the prethermal lifetime $\tau_{\mathrm{pre}}$ is largely independent of the choice of observable whose thermalization is being characterised.

Defining the time $t_x$ via ${I}(t_x)=x$, we extract the prethermal lifetime as the average $\tau_{\mathrm{pre}}=\langle t_x\rangle_x$ where the average over the five values $x=0.8$, $0.8\pm0.015$ and $0.8\pm0.03$ is performed to reduce numerical noise.
Fig.~\ref{fig:dynamics_clean} (c) depicts the dependence of $\tau_{\mathrm{pre}}$ on $1/T$ for different multipolar orders $n=0$, $1$ and $2$.
For $n=1$ and $2$, the numerical results are consistent with the predicted scaling law $\tau_{\mathrm{pre}}\sim T^{-(2n+1)}$~\cite{zhao2021random}. 
For the purely random drive $n=0$, however, 
heating is always fast with only a weak dependence on the driving frequency, suggesting that the multipolar structure provides a qualitative improvement for controlling the intra-band heating~\cite{zhao2021random}.

Apart from the case of purely random driving, $n-$RMD of the single-band model thus allows to increase the prethermal lifetime by choosing a sufficiently large driving frequency $1/T$.
The existence of higher bands in real systems, however, imposes limitations to these choices because fast driving can result in inter-band heating. 

{\it Particle excitation to higher bands.---}
\begin{figure}
	\centering
	\includegraphics[width=0.9\linewidth]{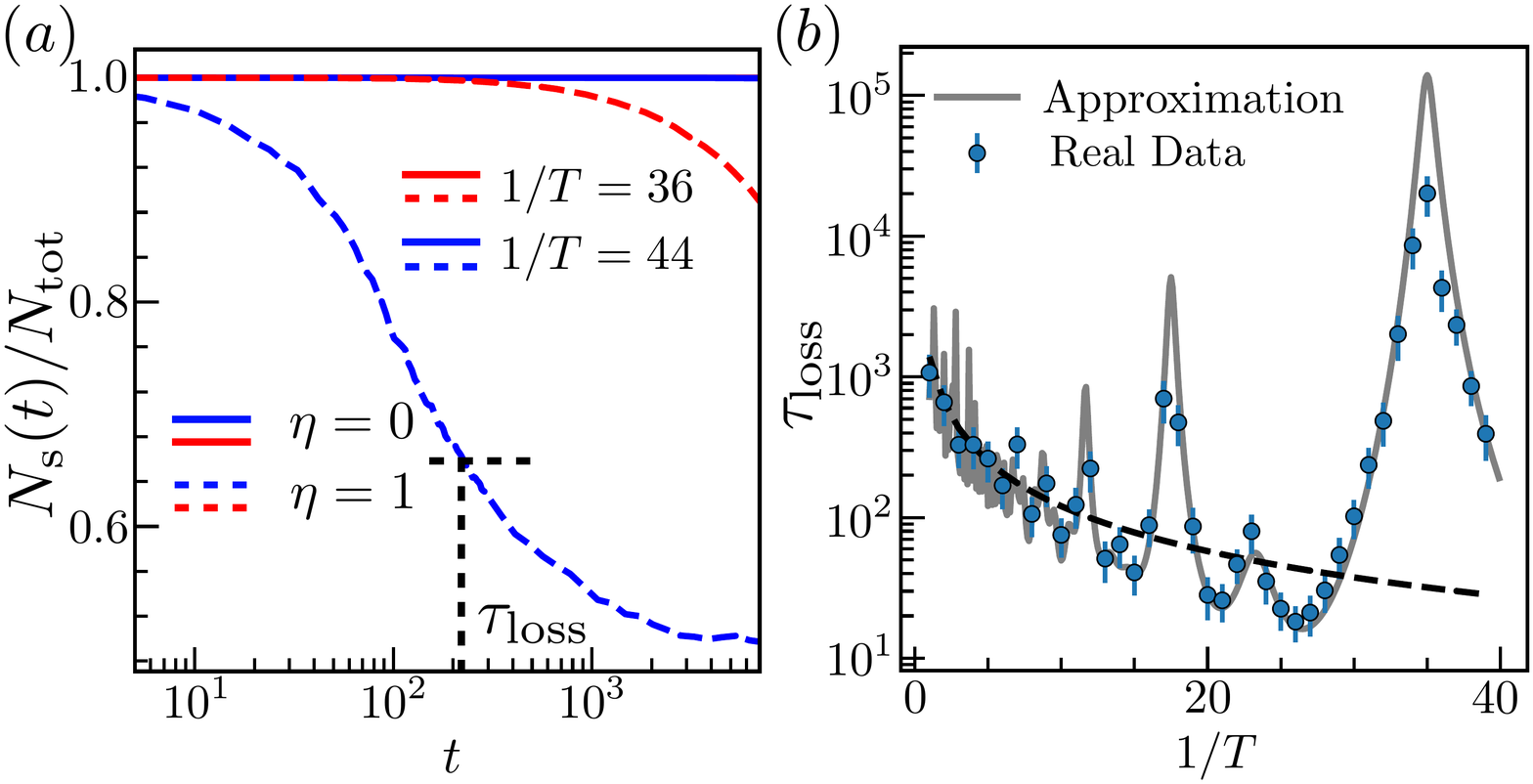}
	\caption{(a) Dynamics of the relative population in s-band for the $2-$RMD. Solid or dashed line corresponds to zero or finite single particle excitations to d-band respectively. Particle loss has a strong dependence on the driving frequency for $\eta=1$. (b) Dependence of the particle loss time scale $\tau_{\mathrm{loss}}$ on the driving frequency $1/T$ for $\eta=1$. Although in general larger $1/T$ induces more particle loss (the $t\sim T$ is depicted as black dashed line to guide the eye), there exist several frequency windows where excitation to d-band is significantly suppressed. We use $\delta J_s=1.2,J_d=7,U_s=5,U_d=3,U_{sd}=2,\mu=10,\Delta=220$ for numerical simulation.}
	\label{fig:Particleloss}
\end{figure}
Both the collective hopping (Eq.~\ref{eq.inter_band}) and the single particle transitions (Eq.~\ref{eq.s-pdrive}) can result in particle excitations to the d-band,
and comparison between dynamics with $\eta=0$ and with $\eta=1$ helps to distinguish between these two processes.
Fig.~\ref{fig:Particleloss} (a) depicts $2-$RMD dynamics of the relative population $N_{\mathrm{s}}/N_{\mathrm{tot}}$ of particles number in the s-band. The initial state is chosen as a density-wave state in the s-band with total particle number $N_{\mathrm{tot}}=4$ and $L=8$ sites.
Solid lines correspond to a system without single-particle inter-band transitions ($\eta=0$), whereas dashed lines correspond to the case with a finite transition ratio $\eta=1$.
Red and blue color indicates the two driving frequencies $1/T=36$ and $44$.
For $\eta=0$ (solid), the relative population remains practically constant for both driving frequencies. It thus suggests that  although the inter-band interaction $U_{sd}$ is nonzero, collective hopping events are rare
due to the low filling factor.
In contrast, finite single-particle transitions (dashed) result in notable particle loss to the d-band, highlighting that this process dominates the inter-band heating. 

It is worth noting that the inter-band heating exhibits a strong dependence on the driving frequency: In the case of the faster of the two drives (blue), the asymptotic equal distribution of particles over both bands is reached around $t\sim10^3$, but for the slower drive (red), at the same time, the population remains almost fully in the s-band. 

The particle loss time $\tau_{\mathrm{loss}}$ at which the relative population $N_{\mathrm{s}}/N_{\mathrm{tot}}$ falls below the threshold value of $0.92$ \footnote{Similarly to above, numerical noise can be suppressed in terms of an average over the threshold values $0.92$, $0.92\pm0.015$ and $0.92\pm0.03$.} 
is depicted versus the frequency $1/T$ with blue dots in Fig.~\ref{fig:Particleloss} (b).
Crucially, the dependence is highly non-monotonic with a series of well pronounced peaks that indicate frequency-regimes in which heating to the d-band is strongly suppressed.
In the relatively slow driving regime ($1/T<10$) the loss time follows the proportionality $\tau_{\mathrm{loss}}\sim T$ (black dashed line),
confirming that faster driving yields stronger inter-band heating~\cite{sun2020optimal}.
While for faster driving ($1/T>10$), the particle loss time $\tau_{\mathrm{loss}}$ oscillates around this proportionality, the amplitude of the oscillations are extremely large. For example, in the frequency window $30<1/T<40$ the actual values of $\tau_{\mathrm{loss}}$ can exceed the proportionality relation by more than two orders of magnitude.
  
  \begin{figure}
	\includegraphics[width=0.9\linewidth]{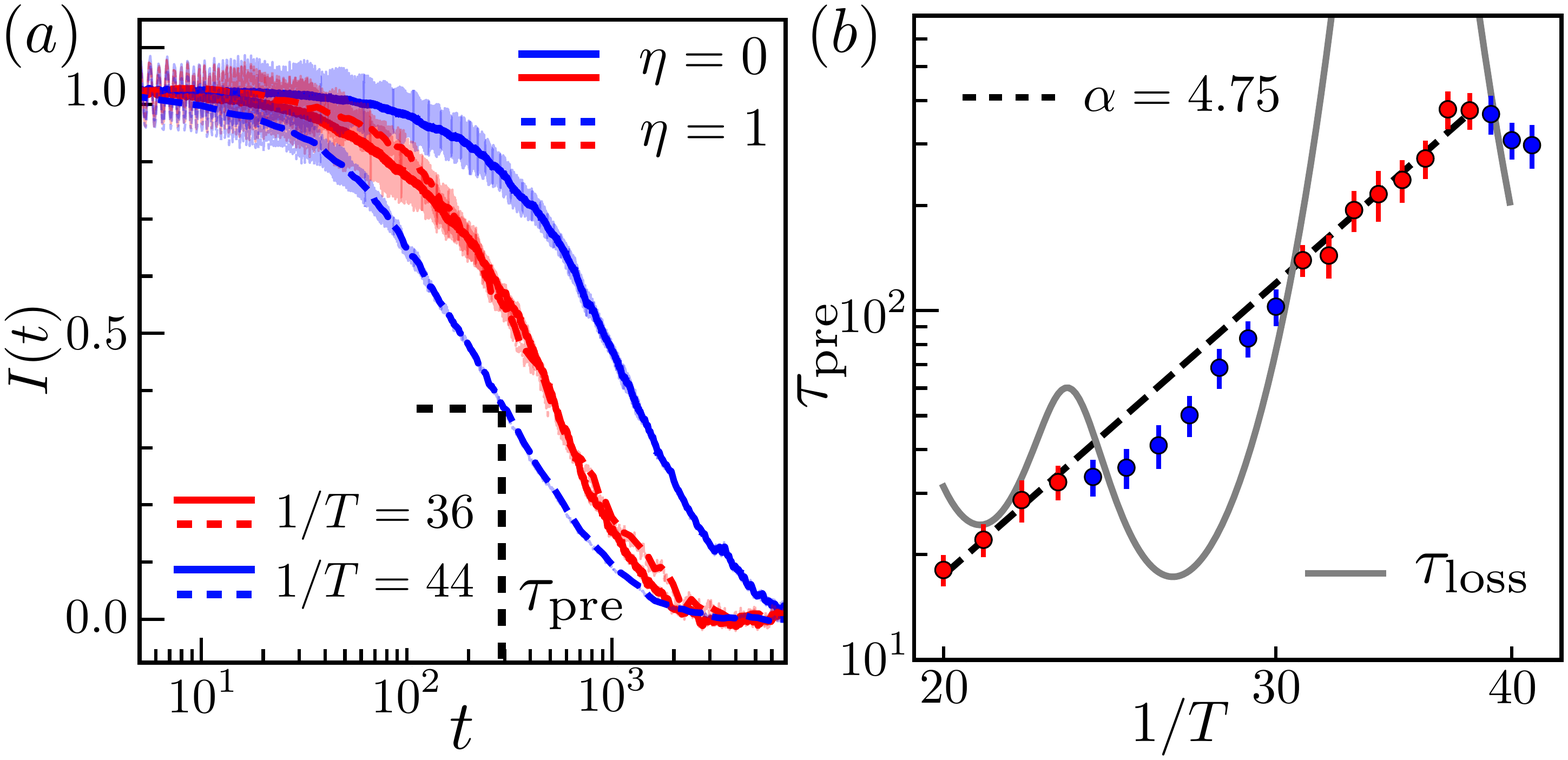}
	\caption{ 
	(a) Dynamics of the imbalance in the s-band for $2-$RMD. For $1/T=44$, the imbalance rapidly decays to zero, whereas for $1/T=36$, the prethermalization remains almost unchanged as notable particle loss only occurs at a later time scale. (b) When particle loss occurs slowly (red dots), the prethermal lifetime  $\tau_{\mathrm{pre}}$ fits well to an algebraic scaling with exponent $\alpha\approx2n+1$. Otherwise, the s-band thermalizes faster (blue dots).}
	\label{fig:Imbalance}
\end{figure}

{\it Tractable model for the heating profile.---}
As the intra-band heating is strongly dominated by single-particle processes, we can provide an analytic approximation to the heating profiles shown in Fig.~\ref{fig:Particleloss} (b).
Neglecting interactions and performing an average over the random choices of time evolution operators in the $n$-RMD protocol yields an exactly solvable model.
Although it is not necessary for solvability, one may further neglect the modulation in the s-band hopping and the staggered potential, as they mainly affect the s-band prethermal dynamics but not the inter-band heating as long as the band gap is  sufficiently large, see Supplementary Material (SM).

Within this approximation, the two-band Hamiltonian can be expressed as  $\hat{H}_{\pm} = \sum_k \boldsymbol{\hat{b}}_k^{\dagger}\hat{H}_k^{\pm}\boldsymbol{\hat{b}}_k$ with the matrix
\begin{equation}
\label{eq.Hamiltonian_k}
\hat{H}_k^{\pm} =\left(\begin{array}{cc}
-2J_s\cos(k) & \pm\delta J_s\eta  \\
\pm\delta J_s\eta & -2J_d\cos(k)+\Delta
\end{array}\right),
\end{equation}
and the vector $\boldsymbol{\hat{b}}_k^{\dagger}=(\hat{b}_{sk}^{\dagger},\hat{b}_{dk}^{\dagger})$ of the creation operator in quasi-momentum space.
The dynamics decomposes into independent quasi-momentum components,
and the elementary time evolution operator in any given component reads
$
U_{0}^{\pm}(k) = \left( C_{0,0}\boldsymbol{1} +i\left(\pm C_{0,1}\hat{\sigma}_x+C_{0,2}\hat{\sigma}_z\right)\right)e^{i\phi_0},
$
with $k$-dependent complex scalars $C_{0,j}$, phases $\phi_0$, the Pauli matrices $\hat{\sigma}_{x,y,z}$ and the identity $\boldsymbol{1}$.
The resulting multipolar operator of $n-$th order reads
$
U_{n}^{\pm}(k)=\left(C_{n, 0}\boldsymbol{1} +i\left(\pm C_{n,1} \hat{\sigma}_{[n]}+C_{n, 2} \hat{\sigma}_{z}\right)\right)e^{i\phi_n},
$
where $[n]=x$ for even $n$ and $[n]=y$ for odd $n$,
and the $k-$dependent coefficients $C_{n,j}$ are determined by the recursive relations
\begin{equation}
\label{eq.C_n}
\begin{aligned}
C_{n+1,0} &= -2C_{n,2}^2+1\ , C_{n+1,2} = 2C_{n,2}C_{n,0}\ , \\
C_{n+1,1} &= -2C_{n,2}C_{n,1}\varepsilon_{z,[n],[n+1]}\ ,
\end{aligned}
\end{equation}
where $\varepsilon_{z,[n],[n+1]}$ denotes the Levi-Civita symbol, see SM.

The time evolution of a single-particle quasi-momentum eigenstate $\ket{\psi_k}=\hat{b}_{sk}^{\dagger}\ket{0}$ is particularly simple but enlightening when $C_{n,1}(k)$ vanishes. This particle oscillates between the two bands, however, as $U_n^{\pm}(k)$ is now diagonal, it completes a full Rabi cycle at the end of each time-window of length $2^nT$.
Therefore, the initial state is recovered and no particle loss happens at stroboscopic times. The dynamics averaged over random choices of the $n$-RMD protocol is still exactly solvable even if we deviate from this special case. To do so,
we use the density matrix $\varrho_k^m$ to represent the state for the $k-$component at time $t=2^nTm$ after the temporal random average~\cite{nielsen2002quantum}. The time evolution can thus be obtained recursively as $
\varrho_k^{m+1}=\frac{1}{2}U_n^+\varrho_k^m (U_n^+)^\dagger+\frac{1}{2}{U}^-_n\varrho^m_k (U^-_n)^\dagger$ with the initial state
$\varrho_k^0 =  \ket{\psi_k}\bra{\psi_k}$.
The resulting density matrix at later times reads
$
\varrho_k^m = f_m\varrho_k^0+g_m\boldsymbol{1} 
$
with coefficients $f_m$ and $g_m$ exactly solvable by induction (see SM).

Even though the dynamics itself does not lead to mixing of  different quasi-momenta in the density matrix,
generic initial states may have such components.
These, however, average out, and the systems' dynamics can be well approximated by the incoherent sum over all momenta. 
If we initialize a Fock state in the s-band, the particle can still complete Rabi-like cycles between two bands at stroboscopic times if the difference between the intra-band hopping rates is much smaller than the band gap ($J_d-J_s\ll\Delta$), see SM.
The resulting population in the s-band is then approximated as
$N_{\mathrm{s}}(2^nTm) \simeq 1/2+\sum_k{(1-2C_{n,1}^2)^m}/{2L},$ suggesting that the population in the s-band decreases exponentially in time.
The particle loss time $\tau_{\mathrm{loss}}$ derived from this model is depicted as a grey line in Fig.~\ref{fig:Particleloss} (b). It matches the exact numerical data with high accuracy and no fitting is needed besides the choice of suitable values of thresholds in the definition of $\tau_{\mathrm{loss}}$.

{\it Protection of prethermalization.---}
Particle loss to higher bands generally prevents the observation of prethermalization in the lowest band. However, the ability to predict regimes of slow inter-band heating can be used to identify suitable prethermalization regimes.
With the driving frequency $1/T=36$, for example, $\tau_{\mathrm{loss}}$ is around $10^4$ (as shown in Fig.~\ref{fig:Particleloss} (b)) which is sufficiently long to make it irrelevant for the prethermal dynamics. Therefore, as shown in Fig.~\ref{fig:Imbalance} (a), there is hardly any noticeable difference between the dynamics with $\eta=0$ (solid) and with $\eta=1$ (dashed) for the red data. But for $1/T=44$ (blue) for which $\tau_{\mathrm{loss}}\sim10^2$ is comparable to the prethermal time scale, heating within the s-band is significantly accelerated by single-particle inter-band transitions. 

The prethermal lifetime $\tau_{\mathrm{pre}}$ defined in terms of the threshold value $0.55$~\footnote{Numerical noise can be suppressed in terms of an average over the threshold values $0.55,0.92\pm0.05$ and $0.92\pm 0.025.$} for the imbalance $I$ in the presence of single particle transition ($\eta=1$) is depicted as function of the driving frequency $1/T$ in  Fig.~\ref{fig:Imbalance} (b) on a log-log scale. 
Apart from the regime of extremely fast driving, as well as the region $25\lesssim 1/T\lesssim 30$, the numerical data agree well with the algebraic scaling $\tau_{\mathrm{pre}}\sim T^{-(2n+1)}$ with $n=2$ indicated by a dashed line.
In particular, in the regime $31\lesssim 1/T\lesssim 39$, the prethermal lifetime in the range $100\lesssim\tau_{\mathrm{pre}}\lesssim400$ is substantially shorter than the particle loss time (grey line). As the accessible lifetime of state-of-the-art quantum simulators is longer than $\tau_{\mathrm{pre}}$~\cite{rubio2020floquet}, we can conclude that the observation of the algebraically long lived prethermal regime~\cite{messer2018floquet, rubio2020floquet,scherg2021observing,yang2020observation} is already within experimental reach.

{\it Discussion.---} 
Inter-band heating remains largely unexplored for aperiodically driven systems, and here we provide the first instance studying the multi-band BHM subjected to the $n-$RMD drive. A surprisingly simple single-particle theory accounts for the non-monotonic inter-band heating profile, which  enables us to identify large frequency windows promising for the experimental observation of the long-lived prethermal plateau and its characteristic lifetime scaling. 

This non-monotonic heating profile is largely independent of the underlying Hamiltonian as long as there is a large gap between the bands. As such, the analytically tractable model should be applicable to other systems. Also, it does not require the driving frequency to be the dominant energy scale as in previous works based on a high frequency expansion~\cite{mori2021rigorous}. Therefore, our work provides a new setting for probing slow thermalization phenomena. It will also be interesting to explore whether the exact solution of the RMD two-level systems can be applied to other contexts, e.g., for integrable many-body systems.

In higher dimensions, lattice geometry plays an important role in suppressing inter-band heating for periodic drives~\cite{messer2018floquet}. It will thus be interesting to explore geometric effects on $n-$RMD drives, the role of dimensionality or the particle density dependence. Reliable numerical simulations of higher dimensional systems normally lie beyond the capability of classical methods especially for random drives. Therefore, experimental investigation of the inter-band heating in quantum simulators would be highly welcome.
 
{\it Acknowledgement.---} We acknowledge very helpful discussions with Monika Aidelsburger and for bringing the  importance of inter-band heating to our attention. We are also grateful for discussions with Bing Yang,
Joseph Vovrosh and Fr\'ed\'eric Sauvage. 
 HZ acknowledges support from a Doctoral-Program Fellowship of the German Academic Exchange Service (DAAD). We acknowledge support from the Imperial-TUM flagship partnership. This work was partly supported by the Deutsche Forschungsgemeinschaft under grants SFB 1143 (project-id 247310070) and the cluster of excellence ct.qmat (EXC 2147, project-id 390858490). The research is part of the Munich Quantum Valley, which is supported by the Bavarian state government with funds from the Hightech Agenda Bayern Plus.
 \bibliography{Heating_bib}

\begin{thebibliography}{49}%
\makeatletter
\providecommand \@ifxundefined [1]{%
 \@ifx{#1\undefined}
}%
\providecommand \@ifnum [1]{%
 \ifnum #1\expandafter \@firstoftwo
 \else \expandafter \@secondoftwo
 \fi
}%
\providecommand \@ifx [1]{%
 \ifx #1\expandafter \@firstoftwo
 \else \expandafter \@secondoftwo
 \fi
}%
\providecommand \natexlab [1]{#1}%
\providecommand \enquote  [1]{``#1''}%
\providecommand \bibnamefont  [1]{#1}%
\providecommand \bibfnamefont [1]{#1}%
\providecommand \citenamefont [1]{#1}%
\providecommand \href@noop [0]{\@secondoftwo}%
\providecommand \href [0]{\begingroup \@sanitize@url \@href}%
\providecommand \@href[1]{\@@startlink{#1}\@@href}%
\providecommand \@@href[1]{\endgroup#1\@@endlink}%
\providecommand \@sanitize@url [0]{\catcode `\\12\catcode `\$12\catcode
  `\&12\catcode `\#12\catcode `\^12\catcode `\_12\catcode `\%12\relax}%
\providecommand \@@startlink[1]{}%
\providecommand \@@endlink[0]{}%
\providecommand \url  [0]{\begingroup\@sanitize@url \@url }%
\providecommand \@url [1]{\endgroup\@href {#1}{\urlprefix }}%
\providecommand \urlprefix  [0]{URL }%
\providecommand \Eprint [0]{\href }%
\providecommand \doibase [0]{http://dx.doi.org/}%
\providecommand \selectlanguage [0]{\@gobble}%
\providecommand \bibinfo  [0]{\@secondoftwo}%
\providecommand \bibfield  [0]{\@secondoftwo}%
\providecommand \translation [1]{[#1]}%
\providecommand \BibitemOpen [0]{}%
\providecommand \bibitemStop [0]{}%
\providecommand \bibitemNoStop [0]{.\EOS\space}%
\providecommand \EOS [0]{\spacefactor3000\relax}%
\providecommand \BibitemShut  [1]{\csname bibitem#1\endcsname}%
\let\auto@bib@innerbib\@empty
\bibitem [{\citenamefont {Moessner}\ and\ \citenamefont
  {Sondhi}(2017)}]{moessner2017equilibration}%
  \BibitemOpen
  \bibfield  {author} {\bibinfo {author} {\bibfnamefont {R.}~\bibnamefont
  {Moessner}}\ and\ \bibinfo {author} {\bibfnamefont {S.~L.}\ \bibnamefont
  {Sondhi}},\ }\href@noop {} {\bibfield  {journal} {\bibinfo  {journal} {Nature
  Physics}\ }\textbf {\bibinfo {volume} {13}},\ \bibinfo {pages} {424}
  (\bibinfo {year} {2017})}\BibitemShut {NoStop}%
\bibitem [{\citenamefont {Wintersperger}\ \emph {et~al.}(2020)\citenamefont
  {Wintersperger}, \citenamefont {Braun}, \citenamefont {{\"U}nal},
  \citenamefont {Eckardt}, \citenamefont {Di~Liberto}, \citenamefont {Goldman},
  \citenamefont {Bloch},\ and\ \citenamefont
  {Aidelsburger}}]{wintersperger2020realization}%
  \BibitemOpen
  \bibfield  {author} {\bibinfo {author} {\bibfnamefont {K.}~\bibnamefont
  {Wintersperger}}, \bibinfo {author} {\bibfnamefont {C.}~\bibnamefont
  {Braun}}, \bibinfo {author} {\bibfnamefont {F.~N.}\ \bibnamefont {{\"U}nal}},
  \bibinfo {author} {\bibfnamefont {A.}~\bibnamefont {Eckardt}}, \bibinfo
  {author} {\bibfnamefont {M.}~\bibnamefont {Di~Liberto}}, \bibinfo {author}
  {\bibfnamefont {N.}~\bibnamefont {Goldman}}, \bibinfo {author} {\bibfnamefont
  {I.}~\bibnamefont {Bloch}}, \ and\ \bibinfo {author} {\bibfnamefont
  {M.}~\bibnamefont {Aidelsburger}},\ }\href@noop {} {\bibfield  {journal}
  {\bibinfo  {journal} {Nature Physics}\ }\textbf {\bibinfo {volume} {16}},\
  \bibinfo {pages} {1058} (\bibinfo {year} {2020})}\BibitemShut {NoStop}%
\bibitem [{\citenamefont {Yang}\ \emph {et~al.}(2020)\citenamefont {Yang},
  \citenamefont {Sun}, \citenamefont {Ott}, \citenamefont {Wang}, \citenamefont
  {Zache}, \citenamefont {Halimeh}, \citenamefont {Yuan}, \citenamefont
  {Hauke},\ and\ \citenamefont {Pan}}]{yang2020observation}%
  \BibitemOpen
  \bibfield  {author} {\bibinfo {author} {\bibfnamefont {B.}~\bibnamefont
  {Yang}}, \bibinfo {author} {\bibfnamefont {H.}~\bibnamefont {Sun}}, \bibinfo
  {author} {\bibfnamefont {R.}~\bibnamefont {Ott}}, \bibinfo {author}
  {\bibfnamefont {H.-Y.}\ \bibnamefont {Wang}}, \bibinfo {author}
  {\bibfnamefont {T.~V.}\ \bibnamefont {Zache}}, \bibinfo {author}
  {\bibfnamefont {J.~C.}\ \bibnamefont {Halimeh}}, \bibinfo {author}
  {\bibfnamefont {Z.-S.}\ \bibnamefont {Yuan}}, \bibinfo {author}
  {\bibfnamefont {P.}~\bibnamefont {Hauke}}, \ and\ \bibinfo {author}
  {\bibfnamefont {J.-W.}\ \bibnamefont {Pan}},\ }\href@noop {} {\bibfield
  {journal} {\bibinfo  {journal} {Nature}\ }\textbf {\bibinfo {volume} {587}},\
  \bibinfo {pages} {392} (\bibinfo {year} {2020})}\BibitemShut {NoStop}%
\bibitem [{\citenamefont {Martinez}\ \emph {et~al.}(2016)\citenamefont
  {Martinez}, \citenamefont {Muschik}, \citenamefont {Schindler}, \citenamefont
  {Nigg}, \citenamefont {Erhard}, \citenamefont {Heyl}, \citenamefont {Hauke},
  \citenamefont {Dalmonte}, \citenamefont {Monz}, \citenamefont {Zoller} \emph
  {et~al.}}]{martinez2016real}%
  \BibitemOpen
  \bibfield  {author} {\bibinfo {author} {\bibfnamefont {E.~A.}\ \bibnamefont
  {Martinez}}, \bibinfo {author} {\bibfnamefont {C.~A.}\ \bibnamefont
  {Muschik}}, \bibinfo {author} {\bibfnamefont {P.}~\bibnamefont {Schindler}},
  \bibinfo {author} {\bibfnamefont {D.}~\bibnamefont {Nigg}}, \bibinfo {author}
  {\bibfnamefont {A.}~\bibnamefont {Erhard}}, \bibinfo {author} {\bibfnamefont
  {M.}~\bibnamefont {Heyl}}, \bibinfo {author} {\bibfnamefont {P.}~\bibnamefont
  {Hauke}}, \bibinfo {author} {\bibfnamefont {M.}~\bibnamefont {Dalmonte}},
  \bibinfo {author} {\bibfnamefont {T.}~\bibnamefont {Monz}}, \bibinfo {author}
  {\bibfnamefont {P.}~\bibnamefont {Zoller}},  \emph {et~al.},\ }\href@noop {}
  {\bibfield  {journal} {\bibinfo  {journal} {Nature}\ }\textbf {\bibinfo
  {volume} {534}},\ \bibinfo {pages} {516} (\bibinfo {year}
  {2016})}\BibitemShut {NoStop}%
\bibitem [{\citenamefont {Arute}\ \emph {et~al.}(2019)\citenamefont {Arute},
  \citenamefont {Arya}, \citenamefont {Babbush}, \citenamefont {Bacon},
  \citenamefont {Bardin}, \citenamefont {Barends}, \citenamefont {Biswas},
  \citenamefont {Boixo}, \citenamefont {Brandao}, \citenamefont {Buell} \emph
  {et~al.}}]{arute2019quantum}%
  \BibitemOpen
  \bibfield  {author} {\bibinfo {author} {\bibfnamefont {F.}~\bibnamefont
  {Arute}}, \bibinfo {author} {\bibfnamefont {K.}~\bibnamefont {Arya}},
  \bibinfo {author} {\bibfnamefont {R.}~\bibnamefont {Babbush}}, \bibinfo
  {author} {\bibfnamefont {D.}~\bibnamefont {Bacon}}, \bibinfo {author}
  {\bibfnamefont {J.~C.}\ \bibnamefont {Bardin}}, \bibinfo {author}
  {\bibfnamefont {R.}~\bibnamefont {Barends}}, \bibinfo {author} {\bibfnamefont
  {R.}~\bibnamefont {Biswas}}, \bibinfo {author} {\bibfnamefont
  {S.}~\bibnamefont {Boixo}}, \bibinfo {author} {\bibfnamefont {F.~G.}\
  \bibnamefont {Brandao}}, \bibinfo {author} {\bibfnamefont {D.~A.}\
  \bibnamefont {Buell}},  \emph {et~al.},\ }\href@noop {} {\bibfield  {journal}
  {\bibinfo  {journal} {Nature}\ }\textbf {\bibinfo {volume} {574}},\ \bibinfo
  {pages} {505} (\bibinfo {year} {2019})}\BibitemShut {NoStop}%
\bibitem [{\citenamefont {Song}\ \emph {et~al.}(2021)\citenamefont {Song},
  \citenamefont {Dutta}, \citenamefont {Bhave}, \citenamefont {Yu},
  \citenamefont {Carter}, \citenamefont {Cooper},\ and\ \citenamefont
  {Schneider}}]{song2021realizing}%
  \BibitemOpen
  \bibfield  {author} {\bibinfo {author} {\bibfnamefont {B.}~\bibnamefont
  {Song}}, \bibinfo {author} {\bibfnamefont {S.}~\bibnamefont {Dutta}},
  \bibinfo {author} {\bibfnamefont {S.}~\bibnamefont {Bhave}}, \bibinfo
  {author} {\bibfnamefont {J.-C.}\ \bibnamefont {Yu}}, \bibinfo {author}
  {\bibfnamefont {E.}~\bibnamefont {Carter}}, \bibinfo {author} {\bibfnamefont
  {N.}~\bibnamefont {Cooper}}, \ and\ \bibinfo {author} {\bibfnamefont
  {U.}~\bibnamefont {Schneider}},\ }\href@noop {} {\bibfield  {journal}
  {\bibinfo  {journal} {arXiv preprint arXiv:2105.12146}\ } (\bibinfo {year}
  {2021})}\BibitemShut {NoStop}%
\bibitem [{\citenamefont {Oka}\ and\ \citenamefont
  {Kitamura}(2019)}]{oka2019floquet}%
  \BibitemOpen
  \bibfield  {author} {\bibinfo {author} {\bibfnamefont {T.}~\bibnamefont
  {Oka}}\ and\ \bibinfo {author} {\bibfnamefont {S.}~\bibnamefont {Kitamura}},\
  }\href@noop {} {\bibfield  {journal} {\bibinfo  {journal} {Annual Review of
  Condensed Matter Physics}\ }\textbf {\bibinfo {volume} {10}},\ \bibinfo
  {pages} {387} (\bibinfo {year} {2019})}\BibitemShut {NoStop}%
\bibitem [{\citenamefont {Bukov}\ \emph {et~al.}(2015)\citenamefont {Bukov},
  \citenamefont {D'Alessio},\ and\ \citenamefont
  {Polkovnikov}}]{bukov2015universal}%
  \BibitemOpen
  \bibfield  {author} {\bibinfo {author} {\bibfnamefont {M.}~\bibnamefont
  {Bukov}}, \bibinfo {author} {\bibfnamefont {L.}~\bibnamefont {D'Alessio}}, \
  and\ \bibinfo {author} {\bibfnamefont {A.}~\bibnamefont {Polkovnikov}},\
  }\href@noop {} {\bibfield  {journal} {\bibinfo  {journal} {Advances in
  Physics}\ }\textbf {\bibinfo {volume} {64}},\ \bibinfo {pages} {139}
  (\bibinfo {year} {2015})}\BibitemShut {NoStop}%
\bibitem [{\citenamefont {Eckardt}(2017)}]{eckardt2017colloquium}%
  \BibitemOpen
  \bibfield  {author} {\bibinfo {author} {\bibfnamefont {A.}~\bibnamefont
  {Eckardt}},\ }\href@noop {} {\bibfield  {journal} {\bibinfo  {journal}
  {Reviews of Modern Physics}\ }\textbf {\bibinfo {volume} {89}},\ \bibinfo
  {pages} {011004} (\bibinfo {year} {2017})}\BibitemShut {NoStop}%
\bibitem [{\citenamefont {Khemani}\ \emph {et~al.}(2016)\citenamefont
  {Khemani}, \citenamefont {Lazarides}, \citenamefont {Moessner},\ and\
  \citenamefont {Sondhi}}]{khemani2016phase}%
  \BibitemOpen
  \bibfield  {author} {\bibinfo {author} {\bibfnamefont {V.}~\bibnamefont
  {Khemani}}, \bibinfo {author} {\bibfnamefont {A.}~\bibnamefont {Lazarides}},
  \bibinfo {author} {\bibfnamefont {R.}~\bibnamefont {Moessner}}, \ and\
  \bibinfo {author} {\bibfnamefont {S.~L.}\ \bibnamefont {Sondhi}},\
  }\href@noop {} {\bibfield  {journal} {\bibinfo  {journal} {Physical review
  letters}\ }\textbf {\bibinfo {volume} {116}},\ \bibinfo {pages} {250401}
  (\bibinfo {year} {2016})}\BibitemShut {NoStop}%
\bibitem [{\citenamefont {Else}\ \emph {et~al.}(2016)\citenamefont {Else},
  \citenamefont {Bauer},\ and\ \citenamefont {Nayak}}]{else2016floquet}%
  \BibitemOpen
  \bibfield  {author} {\bibinfo {author} {\bibfnamefont {D.~V.}\ \bibnamefont
  {Else}}, \bibinfo {author} {\bibfnamefont {B.}~\bibnamefont {Bauer}}, \ and\
  \bibinfo {author} {\bibfnamefont {C.}~\bibnamefont {Nayak}},\ }\href@noop {}
  {\bibfield  {journal} {\bibinfo  {journal} {Physical review letters}\
  }\textbf {\bibinfo {volume} {117}},\ \bibinfo {pages} {090402} (\bibinfo
  {year} {2016})}\BibitemShut {NoStop}%
\bibitem [{\citenamefont {Struck}\ \emph {et~al.}(2013)\citenamefont {Struck},
  \citenamefont {Weinberg}, \citenamefont {{\"O}lschl{\"a}ger}, \citenamefont
  {Windpassinger}, \citenamefont {Simonet}, \citenamefont {Sengstock},
  \citenamefont {H{\"o}ppner}, \citenamefont {Hauke}, \citenamefont {Eckardt},
  \citenamefont {Lewenstein} \emph {et~al.}}]{struck2013engineering}%
  \BibitemOpen
  \bibfield  {author} {\bibinfo {author} {\bibfnamefont {J.}~\bibnamefont
  {Struck}}, \bibinfo {author} {\bibfnamefont {M.}~\bibnamefont {Weinberg}},
  \bibinfo {author} {\bibfnamefont {C.}~\bibnamefont {{\"O}lschl{\"a}ger}},
  \bibinfo {author} {\bibfnamefont {P.}~\bibnamefont {Windpassinger}}, \bibinfo
  {author} {\bibfnamefont {J.}~\bibnamefont {Simonet}}, \bibinfo {author}
  {\bibfnamefont {K.}~\bibnamefont {Sengstock}}, \bibinfo {author}
  {\bibfnamefont {R.}~\bibnamefont {H{\"o}ppner}}, \bibinfo {author}
  {\bibfnamefont {P.}~\bibnamefont {Hauke}}, \bibinfo {author} {\bibfnamefont
  {A.}~\bibnamefont {Eckardt}}, \bibinfo {author} {\bibfnamefont
  {M.}~\bibnamefont {Lewenstein}},  \emph {et~al.},\ }\href@noop {} {\bibfield
  {journal} {\bibinfo  {journal} {Nature Physics}\ }\textbf {\bibinfo {volume}
  {9}},\ \bibinfo {pages} {738} (\bibinfo {year} {2013})}\BibitemShut {NoStop}%
\bibitem [{\citenamefont {Aidelsburger}\ \emph {et~al.}(2013)\citenamefont
  {Aidelsburger}, \citenamefont {Atala}, \citenamefont {Lohse}, \citenamefont
  {Barreiro}, \citenamefont {Paredes},\ and\ \citenamefont
  {Bloch}}]{aidelsburger2013realization}%
  \BibitemOpen
  \bibfield  {author} {\bibinfo {author} {\bibfnamefont {M.}~\bibnamefont
  {Aidelsburger}}, \bibinfo {author} {\bibfnamefont {M.}~\bibnamefont {Atala}},
  \bibinfo {author} {\bibfnamefont {M.}~\bibnamefont {Lohse}}, \bibinfo
  {author} {\bibfnamefont {J.~T.}\ \bibnamefont {Barreiro}}, \bibinfo {author}
  {\bibfnamefont {B.}~\bibnamefont {Paredes}}, \ and\ \bibinfo {author}
  {\bibfnamefont {I.}~\bibnamefont {Bloch}},\ }\href@noop {} {\bibfield
  {journal} {\bibinfo  {journal} {Physical review letters}\ }\textbf {\bibinfo
  {volume} {111}},\ \bibinfo {pages} {185301} (\bibinfo {year}
  {2013})}\BibitemShut {NoStop}%
\bibitem [{\citenamefont {Gross}\ and\ \citenamefont
  {Bloch}(2017)}]{gross2017quantum}%
  \BibitemOpen
  \bibfield  {author} {\bibinfo {author} {\bibfnamefont {C.}~\bibnamefont
  {Gross}}\ and\ \bibinfo {author} {\bibfnamefont {I.}~\bibnamefont {Bloch}},\
  }\href@noop {} {\bibfield  {journal} {\bibinfo  {journal} {Science}\ }\textbf
  {\bibinfo {volume} {357}},\ \bibinfo {pages} {995} (\bibinfo {year}
  {2017})}\BibitemShut {NoStop}%
\bibitem [{\citenamefont {Nathan}\ \emph {et~al.}(2019)\citenamefont {Nathan},
  \citenamefont {Abanin}, \citenamefont {Berg}, \citenamefont {Lindner},\ and\
  \citenamefont {Rudner}}]{nathan2019anomalous}%
  \BibitemOpen
  \bibfield  {author} {\bibinfo {author} {\bibfnamefont {F.}~\bibnamefont
  {Nathan}}, \bibinfo {author} {\bibfnamefont {D.}~\bibnamefont {Abanin}},
  \bibinfo {author} {\bibfnamefont {E.}~\bibnamefont {Berg}}, \bibinfo {author}
  {\bibfnamefont {N.~H.}\ \bibnamefont {Lindner}}, \ and\ \bibinfo {author}
  {\bibfnamefont {M.~S.}\ \bibnamefont {Rudner}},\ }\href@noop {} {\bibfield
  {journal} {\bibinfo  {journal} {Physical Review B}\ }\textbf {\bibinfo
  {volume} {99}},\ \bibinfo {pages} {195133} (\bibinfo {year}
  {2019})}\BibitemShut {NoStop}%
\bibitem [{\citenamefont {Nathan}\ \emph {et~al.}(2017)\citenamefont {Nathan},
  \citenamefont {Rudner}, \citenamefont {Lindner}, \citenamefont {Berg},\ and\
  \citenamefont {Refael}}]{nathan2017quantized}%
  \BibitemOpen
  \bibfield  {author} {\bibinfo {author} {\bibfnamefont {F.}~\bibnamefont
  {Nathan}}, \bibinfo {author} {\bibfnamefont {M.~S.}\ \bibnamefont {Rudner}},
  \bibinfo {author} {\bibfnamefont {N.~H.}\ \bibnamefont {Lindner}}, \bibinfo
  {author} {\bibfnamefont {E.}~\bibnamefont {Berg}}, \ and\ \bibinfo {author}
  {\bibfnamefont {G.}~\bibnamefont {Refael}},\ }\href@noop {} {\bibfield
  {journal} {\bibinfo  {journal} {Physical review letters}\ }\textbf {\bibinfo
  {volume} {119}},\ \bibinfo {pages} {186801} (\bibinfo {year}
  {2017})}\BibitemShut {NoStop}%
\bibitem [{\citenamefont {Kitagawa}\ \emph {et~al.}(2010)\citenamefont
  {Kitagawa}, \citenamefont {Berg}, \citenamefont {Rudner},\ and\ \citenamefont
  {Demler}}]{kitagawa2010topological}%
  \BibitemOpen
  \bibfield  {author} {\bibinfo {author} {\bibfnamefont {T.}~\bibnamefont
  {Kitagawa}}, \bibinfo {author} {\bibfnamefont {E.}~\bibnamefont {Berg}},
  \bibinfo {author} {\bibfnamefont {M.}~\bibnamefont {Rudner}}, \ and\ \bibinfo
  {author} {\bibfnamefont {E.}~\bibnamefont {Demler}},\ }\href@noop {}
  {\bibfield  {journal} {\bibinfo  {journal} {Physical Review B}\ }\textbf
  {\bibinfo {volume} {82}},\ \bibinfo {pages} {235114} (\bibinfo {year}
  {2010})}\BibitemShut {NoStop}%
\bibitem [{\citenamefont {Rubio-Abadal}\ \emph {et~al.}(2020)\citenamefont
  {Rubio-Abadal}, \citenamefont {Ippoliti}, \citenamefont {Hollerith},
  \citenamefont {Wei}, \citenamefont {Rui}, \citenamefont {Sondhi},
  \citenamefont {Khemani}, \citenamefont {Gross},\ and\ \citenamefont
  {Bloch}}]{rubio2020floquet}%
  \BibitemOpen
  \bibfield  {author} {\bibinfo {author} {\bibfnamefont {A.}~\bibnamefont
  {Rubio-Abadal}}, \bibinfo {author} {\bibfnamefont {M.}~\bibnamefont
  {Ippoliti}}, \bibinfo {author} {\bibfnamefont {S.}~\bibnamefont {Hollerith}},
  \bibinfo {author} {\bibfnamefont {D.}~\bibnamefont {Wei}}, \bibinfo {author}
  {\bibfnamefont {J.}~\bibnamefont {Rui}}, \bibinfo {author} {\bibfnamefont
  {S.}~\bibnamefont {Sondhi}}, \bibinfo {author} {\bibfnamefont
  {V.}~\bibnamefont {Khemani}}, \bibinfo {author} {\bibfnamefont
  {C.}~\bibnamefont {Gross}}, \ and\ \bibinfo {author} {\bibfnamefont
  {I.}~\bibnamefont {Bloch}},\ }\href@noop {} {\bibfield  {journal} {\bibinfo
  {journal} {Physical Review X}\ }\textbf {\bibinfo {volume} {10}},\ \bibinfo
  {pages} {021044} (\bibinfo {year} {2020})}\BibitemShut {NoStop}%
\bibitem [{\citenamefont {Reitter}\ \emph {et~al.}(2017)\citenamefont
  {Reitter}, \citenamefont {N{\"a}ger}, \citenamefont {Wintersperger},
  \citenamefont {Str{\"a}ter}, \citenamefont {Bloch}, \citenamefont {Eckardt},\
  and\ \citenamefont {Schneider}}]{reitter2017interaction}%
  \BibitemOpen
  \bibfield  {author} {\bibinfo {author} {\bibfnamefont {M.}~\bibnamefont
  {Reitter}}, \bibinfo {author} {\bibfnamefont {J.}~\bibnamefont {N{\"a}ger}},
  \bibinfo {author} {\bibfnamefont {K.}~\bibnamefont {Wintersperger}}, \bibinfo
  {author} {\bibfnamefont {C.}~\bibnamefont {Str{\"a}ter}}, \bibinfo {author}
  {\bibfnamefont {I.}~\bibnamefont {Bloch}}, \bibinfo {author} {\bibfnamefont
  {A.}~\bibnamefont {Eckardt}}, \ and\ \bibinfo {author} {\bibfnamefont
  {U.}~\bibnamefont {Schneider}},\ }\href@noop {} {\bibfield  {journal}
  {\bibinfo  {journal} {Physical review letters}\ }\textbf {\bibinfo {volume}
  {119}},\ \bibinfo {pages} {200402} (\bibinfo {year} {2017})}\BibitemShut
  {NoStop}%
\bibitem [{\citenamefont {Viebahn}\ \emph {et~al.}(2021)\citenamefont
  {Viebahn}, \citenamefont {Minguzzi}, \citenamefont {Sandholzer},
  \citenamefont {Walter}, \citenamefont {Sajnani}, \citenamefont {G{\"o}rg},\
  and\ \citenamefont {Esslinger}}]{viebahn2021suppressing}%
  \BibitemOpen
  \bibfield  {author} {\bibinfo {author} {\bibfnamefont {K.}~\bibnamefont
  {Viebahn}}, \bibinfo {author} {\bibfnamefont {J.}~\bibnamefont {Minguzzi}},
  \bibinfo {author} {\bibfnamefont {K.}~\bibnamefont {Sandholzer}}, \bibinfo
  {author} {\bibfnamefont {A.-S.}\ \bibnamefont {Walter}}, \bibinfo {author}
  {\bibfnamefont {M.}~\bibnamefont {Sajnani}}, \bibinfo {author} {\bibfnamefont
  {F.}~\bibnamefont {G{\"o}rg}}, \ and\ \bibinfo {author} {\bibfnamefont
  {T.}~\bibnamefont {Esslinger}},\ }\href@noop {} {\bibfield  {journal}
  {\bibinfo  {journal} {Physical Review X}\ }\textbf {\bibinfo {volume} {11}},\
  \bibinfo {pages} {011057} (\bibinfo {year} {2021})}\BibitemShut {NoStop}%
\bibitem [{\citenamefont {Lazarides}\ \emph {et~al.}(2014)\citenamefont
  {Lazarides}, \citenamefont {Das},\ and\ \citenamefont
  {Moessner}}]{lazarides2014equilibrium}%
  \BibitemOpen
  \bibfield  {author} {\bibinfo {author} {\bibfnamefont {A.}~\bibnamefont
  {Lazarides}}, \bibinfo {author} {\bibfnamefont {A.}~\bibnamefont {Das}}, \
  and\ \bibinfo {author} {\bibfnamefont {R.}~\bibnamefont {Moessner}},\
  }\href@noop {} {\bibfield  {journal} {\bibinfo  {journal} {Physical Review
  E}\ }\textbf {\bibinfo {volume} {90}},\ \bibinfo {pages} {012110} (\bibinfo
  {year} {2014})}\BibitemShut {NoStop}%
\bibitem [{\citenamefont {Abanin}\ \emph {et~al.}(2015)\citenamefont {Abanin},
  \citenamefont {De~Roeck},\ and\ \citenamefont
  {Huveneers}}]{abanin2015exponentially}%
  \BibitemOpen
  \bibfield  {author} {\bibinfo {author} {\bibfnamefont {D.~A.}\ \bibnamefont
  {Abanin}}, \bibinfo {author} {\bibfnamefont {W.}~\bibnamefont {De~Roeck}}, \
  and\ \bibinfo {author} {\bibfnamefont {F.}~\bibnamefont {Huveneers}},\
  }\href@noop {} {\bibfield  {journal} {\bibinfo  {journal} {Physical review
  letters}\ }\textbf {\bibinfo {volume} {115}},\ \bibinfo {pages} {256803}
  (\bibinfo {year} {2015})}\BibitemShut {NoStop}%
\bibitem [{\citenamefont {Mori}\ \emph {et~al.}(2018)\citenamefont {Mori},
  \citenamefont {Ikeda}, \citenamefont {Kaminishi},\ and\ \citenamefont
  {Ueda}}]{mori2018thermalization}%
  \BibitemOpen
  \bibfield  {author} {\bibinfo {author} {\bibfnamefont {T.}~\bibnamefont
  {Mori}}, \bibinfo {author} {\bibfnamefont {T.~N.}\ \bibnamefont {Ikeda}},
  \bibinfo {author} {\bibfnamefont {E.}~\bibnamefont {Kaminishi}}, \ and\
  \bibinfo {author} {\bibfnamefont {M.}~\bibnamefont {Ueda}},\ }\href@noop {}
  {\bibfield  {journal} {\bibinfo  {journal} {Journal of Physics B: Atomic,
  Molecular and Optical Physics}\ }\textbf {\bibinfo {volume} {51}},\ \bibinfo
  {pages} {112001} (\bibinfo {year} {2018})}\BibitemShut {NoStop}%
\bibitem [{\citenamefont {Kuwahara}\ \emph {et~al.}(2016)\citenamefont
  {Kuwahara}, \citenamefont {Mori},\ and\ \citenamefont
  {Saito}}]{kuwahara2016floquet}%
  \BibitemOpen
  \bibfield  {author} {\bibinfo {author} {\bibfnamefont {T.}~\bibnamefont
  {Kuwahara}}, \bibinfo {author} {\bibfnamefont {T.}~\bibnamefont {Mori}}, \
  and\ \bibinfo {author} {\bibfnamefont {K.}~\bibnamefont {Saito}},\
  }\href@noop {} {\bibfield  {journal} {\bibinfo  {journal} {Annals of
  Physics}\ }\textbf {\bibinfo {volume} {367}},\ \bibinfo {pages} {96}
  (\bibinfo {year} {2016})}\BibitemShut {NoStop}%
\bibitem [{\citenamefont {Abanin}\ \emph {et~al.}(2017)\citenamefont {Abanin},
  \citenamefont {De~Roeck}, \citenamefont {Ho},\ and\ \citenamefont
  {Huveneers}}]{abanin2017rigorous}%
  \BibitemOpen
  \bibfield  {author} {\bibinfo {author} {\bibfnamefont {D.}~\bibnamefont
  {Abanin}}, \bibinfo {author} {\bibfnamefont {W.}~\bibnamefont {De~Roeck}},
  \bibinfo {author} {\bibfnamefont {W.~W.}\ \bibnamefont {Ho}}, \ and\ \bibinfo
  {author} {\bibfnamefont {F.}~\bibnamefont {Huveneers}},\ }\href@noop {}
  {\bibfield  {journal} {\bibinfo  {journal} {Communications in Mathematical
  Physics}\ }\textbf {\bibinfo {volume} {354}},\ \bibinfo {pages} {809}
  (\bibinfo {year} {2017})}\BibitemShut {NoStop}%
\bibitem [{\citenamefont {Eckardt}\ and\ \citenamefont
  {Anisimovas}(2015)}]{eckardt2015high}%
  \BibitemOpen
  \bibfield  {author} {\bibinfo {author} {\bibfnamefont {A.}~\bibnamefont
  {Eckardt}}\ and\ \bibinfo {author} {\bibfnamefont {E.}~\bibnamefont
  {Anisimovas}},\ }\href@noop {} {\bibfield  {journal} {\bibinfo  {journal}
  {New journal of physics}\ }\textbf {\bibinfo {volume} {17}},\ \bibinfo
  {pages} {093039} (\bibinfo {year} {2015})}\BibitemShut {NoStop}%
\bibitem [{\citenamefont {Machado}\ \emph {et~al.}(2020)\citenamefont
  {Machado}, \citenamefont {Else}, \citenamefont {Kahanamoku-Meyer},
  \citenamefont {Nayak},\ and\ \citenamefont {Yao}}]{machado2020long}%
  \BibitemOpen
  \bibfield  {author} {\bibinfo {author} {\bibfnamefont {F.}~\bibnamefont
  {Machado}}, \bibinfo {author} {\bibfnamefont {D.~V.}\ \bibnamefont {Else}},
  \bibinfo {author} {\bibfnamefont {G.~D.}\ \bibnamefont {Kahanamoku-Meyer}},
  \bibinfo {author} {\bibfnamefont {C.}~\bibnamefont {Nayak}}, \ and\ \bibinfo
  {author} {\bibfnamefont {N.~Y.}\ \bibnamefont {Yao}},\ }\href@noop {}
  {\bibfield  {journal} {\bibinfo  {journal} {Physical Review X}\ }\textbf
  {\bibinfo {volume} {10}},\ \bibinfo {pages} {011043} (\bibinfo {year}
  {2020})}\BibitemShut {NoStop}%
\bibitem [{\citenamefont {Pizzi}\ \emph {et~al.}(2021)\citenamefont {Pizzi},
  \citenamefont {Nunnenkamp},\ and\ \citenamefont
  {Knolle}}]{pizzi2021classical}%
  \BibitemOpen
  \bibfield  {author} {\bibinfo {author} {\bibfnamefont {A.}~\bibnamefont
  {Pizzi}}, \bibinfo {author} {\bibfnamefont {A.}~\bibnamefont {Nunnenkamp}}, \
  and\ \bibinfo {author} {\bibfnamefont {J.}~\bibnamefont {Knolle}},\
  }\href@noop {} {\bibfield  {journal} {\bibinfo  {journal} {arXiv preprint
  arXiv:2104.13928}\ } (\bibinfo {year} {2021})}\BibitemShut {NoStop}%
\bibitem [{\citenamefont {Sun}\ and\ \citenamefont
  {Eckardt}(2020)}]{sun2020optimal}%
  \BibitemOpen
  \bibfield  {author} {\bibinfo {author} {\bibfnamefont {G.}~\bibnamefont
  {Sun}}\ and\ \bibinfo {author} {\bibfnamefont {A.}~\bibnamefont {Eckardt}},\
  }\href@noop {} {\bibfield  {journal} {\bibinfo  {journal} {Physical Review
  Research}\ }\textbf {\bibinfo {volume} {2}},\ \bibinfo {pages} {013241}
  (\bibinfo {year} {2020})}\BibitemShut {NoStop}%
\bibitem [{\citenamefont {Beatrez}\ \emph {et~al.}(2021)\citenamefont
  {Beatrez}, \citenamefont {Janes}, \citenamefont {Akkiraju}, \citenamefont
  {Pillai}, \citenamefont {Oddo}, \citenamefont {Reshetikhin}, \citenamefont
  {Druga}, \citenamefont {McAllister}, \citenamefont {Elo}, \citenamefont
  {Gilbert} \emph {et~al.}}]{beatrez2021floquet}%
  \BibitemOpen
  \bibfield  {author} {\bibinfo {author} {\bibfnamefont {W.}~\bibnamefont
  {Beatrez}}, \bibinfo {author} {\bibfnamefont {O.}~\bibnamefont {Janes}},
  \bibinfo {author} {\bibfnamefont {A.}~\bibnamefont {Akkiraju}}, \bibinfo
  {author} {\bibfnamefont {A.}~\bibnamefont {Pillai}}, \bibinfo {author}
  {\bibfnamefont {A.}~\bibnamefont {Oddo}}, \bibinfo {author} {\bibfnamefont
  {P.}~\bibnamefont {Reshetikhin}}, \bibinfo {author} {\bibfnamefont
  {E.}~\bibnamefont {Druga}}, \bibinfo {author} {\bibfnamefont
  {M.}~\bibnamefont {McAllister}}, \bibinfo {author} {\bibfnamefont
  {M.}~\bibnamefont {Elo}}, \bibinfo {author} {\bibfnamefont {B.}~\bibnamefont
  {Gilbert}},  \emph {et~al.},\ }\href@noop {} {\bibfield  {journal} {\bibinfo
  {journal} {arXiv preprint arXiv:2104.01988}\ } (\bibinfo {year}
  {2021})}\BibitemShut {NoStop}%
\bibitem [{\citenamefont {Kyprianidis}\ \emph {et~al.}(2021)\citenamefont
  {Kyprianidis}, \citenamefont {Machado}, \citenamefont {Morong}, \citenamefont
  {Becker}, \citenamefont {Collins}, \citenamefont {Else}, \citenamefont
  {Feng}, \citenamefont {Hess}, \citenamefont {Nayak}, \citenamefont {Pagano}
  \emph {et~al.}}]{kyprianidis2021observation}%
  \BibitemOpen
  \bibfield  {author} {\bibinfo {author} {\bibfnamefont {A.}~\bibnamefont
  {Kyprianidis}}, \bibinfo {author} {\bibfnamefont {F.}~\bibnamefont
  {Machado}}, \bibinfo {author} {\bibfnamefont {W.}~\bibnamefont {Morong}},
  \bibinfo {author} {\bibfnamefont {P.}~\bibnamefont {Becker}}, \bibinfo
  {author} {\bibfnamefont {K.~S.}\ \bibnamefont {Collins}}, \bibinfo {author}
  {\bibfnamefont {D.~V.}\ \bibnamefont {Else}}, \bibinfo {author}
  {\bibfnamefont {L.}~\bibnamefont {Feng}}, \bibinfo {author} {\bibfnamefont
  {P.~W.}\ \bibnamefont {Hess}}, \bibinfo {author} {\bibfnamefont
  {C.}~\bibnamefont {Nayak}}, \bibinfo {author} {\bibfnamefont
  {G.}~\bibnamefont {Pagano}},  \emph {et~al.},\ }\href@noop {} {\bibfield
  {journal} {\bibinfo  {journal} {Science}\ }\textbf {\bibinfo {volume}
  {372}},\ \bibinfo {pages} {1192} (\bibinfo {year} {2021})}\BibitemShut
  {NoStop}%
\bibitem [{\citenamefont {Peng}\ \emph {et~al.}(2021)\citenamefont {Peng},
  \citenamefont {Yin}, \citenamefont {Huang}, \citenamefont {Ramanathan},\ and\
  \citenamefont {Cappellaro}}]{peng2021floquet}%
  \BibitemOpen
  \bibfield  {author} {\bibinfo {author} {\bibfnamefont {P.}~\bibnamefont
  {Peng}}, \bibinfo {author} {\bibfnamefont {C.}~\bibnamefont {Yin}}, \bibinfo
  {author} {\bibfnamefont {X.}~\bibnamefont {Huang}}, \bibinfo {author}
  {\bibfnamefont {C.}~\bibnamefont {Ramanathan}}, \ and\ \bibinfo {author}
  {\bibfnamefont {P.}~\bibnamefont {Cappellaro}},\ }\href@noop {} {\bibfield
  {journal} {\bibinfo  {journal} {Nature Physics}\ }\textbf {\bibinfo {volume}
  {17}},\ \bibinfo {pages} {444} (\bibinfo {year} {2021})}\BibitemShut
  {NoStop}%
\bibitem [{\citenamefont {Martin}\ \emph {et~al.}(2017)\citenamefont {Martin},
  \citenamefont {Refael},\ and\ \citenamefont
  {Halperin}}]{martin2017topological}%
  \BibitemOpen
  \bibfield  {author} {\bibinfo {author} {\bibfnamefont {I.}~\bibnamefont
  {Martin}}, \bibinfo {author} {\bibfnamefont {G.}~\bibnamefont {Refael}}, \
  and\ \bibinfo {author} {\bibfnamefont {B.}~\bibnamefont {Halperin}},\
  }\href@noop {} {\bibfield  {journal} {\bibinfo  {journal} {Physical Review
  X}\ }\textbf {\bibinfo {volume} {7}},\ \bibinfo {pages} {041008} (\bibinfo
  {year} {2017})}\BibitemShut {NoStop}%
\bibitem [{\citenamefont {Long}\ \emph {et~al.}(2021)\citenamefont {Long},
  \citenamefont {Crowley},\ and\ \citenamefont {Chandran}}]{long2021many}%
  \BibitemOpen
  \bibfield  {author} {\bibinfo {author} {\bibfnamefont {D.~M.}\ \bibnamefont
  {Long}}, \bibinfo {author} {\bibfnamefont {P.~J.}\ \bibnamefont {Crowley}}, \
  and\ \bibinfo {author} {\bibfnamefont {A.}~\bibnamefont {Chandran}},\
  }\href@noop {} {\bibfield  {journal} {\bibinfo  {journal} {arXiv preprint
  arXiv:2108.04834}\ } (\bibinfo {year} {2021})}\BibitemShut {NoStop}%
\bibitem [{\citenamefont {Friedman}\ \emph {et~al.}(2020)\citenamefont
  {Friedman}, \citenamefont {Ware}, \citenamefont {Vasseur},\ and\
  \citenamefont {Potter}}]{friedman2020topological}%
  \BibitemOpen
  \bibfield  {author} {\bibinfo {author} {\bibfnamefont {A.~J.}\ \bibnamefont
  {Friedman}}, \bibinfo {author} {\bibfnamefont {B.}~\bibnamefont {Ware}},
  \bibinfo {author} {\bibfnamefont {R.}~\bibnamefont {Vasseur}}, \ and\
  \bibinfo {author} {\bibfnamefont {A.~C.}\ \bibnamefont {Potter}},\
  }\href@noop {} {\bibfield  {journal} {\bibinfo  {journal} {arXiv preprint
  arXiv:2009.03314}\ } (\bibinfo {year} {2020})}\BibitemShut {NoStop}%
\bibitem [{\citenamefont {Zhao}\ \emph {et~al.}(2021)\citenamefont {Zhao},
  \citenamefont {Mintert}, \citenamefont {Moessner},\ and\ \citenamefont
  {Knolle}}]{zhao2021random}%
  \BibitemOpen
  \bibfield  {author} {\bibinfo {author} {\bibfnamefont {H.}~\bibnamefont
  {Zhao}}, \bibinfo {author} {\bibfnamefont {F.}~\bibnamefont {Mintert}},
  \bibinfo {author} {\bibfnamefont {R.}~\bibnamefont {Moessner}}, \ and\
  \bibinfo {author} {\bibfnamefont {J.}~\bibnamefont {Knolle}},\ }\href@noop {}
  {\bibfield  {journal} {\bibinfo  {journal} {Physical Review Letters}\
  }\textbf {\bibinfo {volume} {126}},\ \bibinfo {pages} {040601} (\bibinfo
  {year} {2021})}\BibitemShut {NoStop}%
\bibitem [{\citenamefont {Nandy}\ \emph {et~al.}(2017)\citenamefont {Nandy},
  \citenamefont {Sen},\ and\ \citenamefont {Sen}}]{nandy2017aperiodically}%
  \BibitemOpen
  \bibfield  {author} {\bibinfo {author} {\bibfnamefont {S.}~\bibnamefont
  {Nandy}}, \bibinfo {author} {\bibfnamefont {A.}~\bibnamefont {Sen}}, \ and\
  \bibinfo {author} {\bibfnamefont {D.}~\bibnamefont {Sen}},\ }\href@noop {}
  {\bibfield  {journal} {\bibinfo  {journal} {Physical Review X}\ }\textbf
  {\bibinfo {volume} {7}},\ \bibinfo {pages} {031034} (\bibinfo {year}
  {2017})}\BibitemShut {NoStop}%
\bibitem [{\citenamefont {Crowley}\ \emph {et~al.}(2019)\citenamefont
  {Crowley}, \citenamefont {Martin},\ and\ \citenamefont
  {Chandran}}]{crowley2019topological}%
  \BibitemOpen
  \bibfield  {author} {\bibinfo {author} {\bibfnamefont {P.~J.}\ \bibnamefont
  {Crowley}}, \bibinfo {author} {\bibfnamefont {I.}~\bibnamefont {Martin}}, \
  and\ \bibinfo {author} {\bibfnamefont {A.}~\bibnamefont {Chandran}},\
  }\href@noop {} {\bibfield  {journal} {\bibinfo  {journal} {Physical Review
  B}\ }\textbf {\bibinfo {volume} {99}},\ \bibinfo {pages} {064306} (\bibinfo
  {year} {2019})}\BibitemShut {NoStop}%
\bibitem [{\citenamefont {Nathan}\ \emph {et~al.}(2020)\citenamefont {Nathan},
  \citenamefont {Ge}, \citenamefont {Gazit}, \citenamefont {Rudner},\ and\
  \citenamefont {Kolodrubetz}}]{nathan2020quasiperiodic}%
  \BibitemOpen
  \bibfield  {author} {\bibinfo {author} {\bibfnamefont {F.}~\bibnamefont
  {Nathan}}, \bibinfo {author} {\bibfnamefont {R.}~\bibnamefont {Ge}}, \bibinfo
  {author} {\bibfnamefont {S.}~\bibnamefont {Gazit}}, \bibinfo {author}
  {\bibfnamefont {M.~S.}\ \bibnamefont {Rudner}}, \ and\ \bibinfo {author}
  {\bibfnamefont {M.}~\bibnamefont {Kolodrubetz}},\ }\href@noop {} {\bibfield
  {journal} {\bibinfo  {journal} {arXiv preprint arXiv:2010.11485}\ } (\bibinfo
  {year} {2020})}\BibitemShut {NoStop}%
\bibitem [{\citenamefont {Malz}\ and\ \citenamefont
  {Smith}(2021)}]{malz2021topological}%
  \BibitemOpen
  \bibfield  {author} {\bibinfo {author} {\bibfnamefont {D.}~\bibnamefont
  {Malz}}\ and\ \bibinfo {author} {\bibfnamefont {A.}~\bibnamefont {Smith}},\
  }\href@noop {} {\bibfield  {journal} {\bibinfo  {journal} {Physical Review
  Letters}\ }\textbf {\bibinfo {volume} {126}},\ \bibinfo {pages} {163602}
  (\bibinfo {year} {2021})}\BibitemShut {NoStop}%
\bibitem [{\citenamefont {Dumitrescu}\ \emph {et~al.}(2018)\citenamefont
  {Dumitrescu}, \citenamefont {Vasseur},\ and\ \citenamefont
  {Potter}}]{dumitrescu2018logarithmically}%
  \BibitemOpen
  \bibfield  {author} {\bibinfo {author} {\bibfnamefont {P.~T.}\ \bibnamefont
  {Dumitrescu}}, \bibinfo {author} {\bibfnamefont {R.}~\bibnamefont {Vasseur}},
  \ and\ \bibinfo {author} {\bibfnamefont {A.~C.}\ \bibnamefont {Potter}},\
  }\href@noop {} {\bibfield  {journal} {\bibinfo  {journal} {Physical review
  letters}\ }\textbf {\bibinfo {volume} {120}},\ \bibinfo {pages} {070602}
  (\bibinfo {year} {2018})}\BibitemShut {NoStop}%
\bibitem [{\citenamefont {Zhao}\ \emph {et~al.}(2019)\citenamefont {Zhao},
  \citenamefont {Mintert},\ and\ \citenamefont {Knolle}}]{zhao2019floquet}%
  \BibitemOpen
  \bibfield  {author} {\bibinfo {author} {\bibfnamefont {H.}~\bibnamefont
  {Zhao}}, \bibinfo {author} {\bibfnamefont {F.}~\bibnamefont {Mintert}}, \
  and\ \bibinfo {author} {\bibfnamefont {J.}~\bibnamefont {Knolle}},\
  }\href@noop {} {\bibfield  {journal} {\bibinfo  {journal} {Physical Review
  B}\ }\textbf {\bibinfo {volume} {100}},\ \bibinfo {pages} {134302} (\bibinfo
  {year} {2019})}\BibitemShut {NoStop}%
\bibitem [{\citenamefont {Else}\ \emph {et~al.}(2020)\citenamefont {Else},
  \citenamefont {Ho},\ and\ \citenamefont {Dumitrescu}}]{else2020long}%
  \BibitemOpen
  \bibfield  {author} {\bibinfo {author} {\bibfnamefont {D.~V.}\ \bibnamefont
  {Else}}, \bibinfo {author} {\bibfnamefont {W.~W.}\ \bibnamefont {Ho}}, \ and\
  \bibinfo {author} {\bibfnamefont {P.~T.}\ \bibnamefont {Dumitrescu}},\
  }\href@noop {} {\bibfield  {journal} {\bibinfo  {journal} {Physical Review
  X}\ }\textbf {\bibinfo {volume} {10}},\ \bibinfo {pages} {021032} (\bibinfo
  {year} {2020})}\BibitemShut {NoStop}%
\bibitem [{\citenamefont {Mori}\ \emph {et~al.}(2021)\citenamefont {Mori},
  \citenamefont {Zhao}, \citenamefont {Mintert}, \citenamefont {Knolle},\ and\
  \citenamefont {Moessner}}]{mori2021rigorous}%
  \BibitemOpen
  \bibfield  {author} {\bibinfo {author} {\bibfnamefont {T.}~\bibnamefont
  {Mori}}, \bibinfo {author} {\bibfnamefont {H.}~\bibnamefont {Zhao}}, \bibinfo
  {author} {\bibfnamefont {F.}~\bibnamefont {Mintert}}, \bibinfo {author}
  {\bibfnamefont {J.}~\bibnamefont {Knolle}}, \ and\ \bibinfo {author}
  {\bibfnamefont {R.}~\bibnamefont {Moessner}},\ }\href@noop {} {\bibfield
  {journal} {\bibinfo  {journal} {arXiv preprint arXiv:2101.07065}\ } (\bibinfo
  {year} {2021})}\BibitemShut {NoStop}%
\bibitem [{Note1()}]{Note1}%
  \BibitemOpen
  \bibinfo {note} {Similarly to above, numerical noise can be suppressed in
  terms of an average over the threshold values $0.92$, $0.92\pm 0.015$ and
  $0.92\pm 0.03$.}\BibitemShut {Stop}%
\bibitem [{\citenamefont {Nielsen}\ and\ \citenamefont
  {Chuang}(2002)}]{nielsen2002quantum}%
  \BibitemOpen
  \bibfield  {author} {\bibinfo {author} {\bibfnamefont {M.~A.}\ \bibnamefont
  {Nielsen}}\ and\ \bibinfo {author} {\bibfnamefont {I.}~\bibnamefont
  {Chuang}},\ }\href@noop {} {\enquote {\bibinfo {title} {Quantum computation
  and quantum information},}\ } (\bibinfo {year} {2002})\BibitemShut {NoStop}%
\bibitem [{Note2()}]{Note2}%
  \BibitemOpen
  \bibinfo {note} {Numerical noise can be suppressed in terms of an average
  over the threshold values $0.55,0.92\pm 0.05$ and $0.92\pm
  0.025.$}\BibitemShut {NoStop}%
\bibitem [{\citenamefont {Messer}\ \emph {et~al.}(2018)\citenamefont {Messer},
  \citenamefont {Sandholzer}, \citenamefont {G{\"o}rg}, \citenamefont
  {Minguzzi}, \citenamefont {Desbuquois},\ and\ \citenamefont
  {Esslinger}}]{messer2018floquet}%
  \BibitemOpen
  \bibfield  {author} {\bibinfo {author} {\bibfnamefont {M.}~\bibnamefont
  {Messer}}, \bibinfo {author} {\bibfnamefont {K.}~\bibnamefont {Sandholzer}},
  \bibinfo {author} {\bibfnamefont {F.}~\bibnamefont {G{\"o}rg}}, \bibinfo
  {author} {\bibfnamefont {J.}~\bibnamefont {Minguzzi}}, \bibinfo {author}
  {\bibfnamefont {R.}~\bibnamefont {Desbuquois}}, \ and\ \bibinfo {author}
  {\bibfnamefont {T.}~\bibnamefont {Esslinger}},\ }\href@noop {} {\bibfield
  {journal} {\bibinfo  {journal} {Physical review letters}\ }\textbf {\bibinfo
  {volume} {121}},\ \bibinfo {pages} {233603} (\bibinfo {year}
  {2018})}\BibitemShut {NoStop}%
\bibitem [{\citenamefont {Scherg}\ \emph {et~al.}(2021)\citenamefont {Scherg},
  \citenamefont {Kohlert}, \citenamefont {Sala}, \citenamefont {Pollmann},
  \citenamefont {Madhusudhana}, \citenamefont {Bloch},\ and\ \citenamefont
  {Aidelsburger}}]{scherg2021observing}%
  \BibitemOpen
  \bibfield  {author} {\bibinfo {author} {\bibfnamefont {S.}~\bibnamefont
  {Scherg}}, \bibinfo {author} {\bibfnamefont {T.}~\bibnamefont {Kohlert}},
  \bibinfo {author} {\bibfnamefont {P.}~\bibnamefont {Sala}}, \bibinfo {author}
  {\bibfnamefont {F.}~\bibnamefont {Pollmann}}, \bibinfo {author}
  {\bibfnamefont {B.~H.}\ \bibnamefont {Madhusudhana}}, \bibinfo {author}
  {\bibfnamefont {I.}~\bibnamefont {Bloch}}, \ and\ \bibinfo {author}
  {\bibfnamefont {M.}~\bibnamefont {Aidelsburger}},\ }\href@noop {} {\bibfield
  {journal} {\bibinfo  {journal} {Nature Communications}\ }\textbf {\bibinfo
  {volume} {12}},\ \bibinfo {pages} {1} (\bibinfo {year} {2021})}\BibitemShut
  {NoStop}%
\end{thebibliography}%
 
\newpage
\appendix
\section{Details of the analytical calculations}
\paragraph{$k-$subspace Hamiltonian.--} Here, we only consider the single-particle hopping process within the $s$ and $d-$bands, and single-particle transition between the two.
In each $k-$space, the Hamiltonian reduces to
\begin{equation}
\label{eq.Hamiltonian_k}
\hat{H}_k^{\pm} =\left(\begin{array}{cc}
-2J_s\cos(k) & \pm\delta J_s\eta  \\
\pm\delta J_s\eta & -2J_d\cos(k)+\Delta
\end{array}\right).
\end{equation}
In terms of Pauli matrices, we have
\begin{eqnarray}
\begin{aligned}
\hat{H}_k^{\pm}&= \left[\cos(k)(-J_s+J_d)-\frac{\Delta}{2}\right]\sigma_z \pm\delta J_s\eta\sigma_x \\
&+\left[(-J_s-J_d)\cos(k)+\frac{\Delta}{2}\right]\boldsymbol{1},
\end{aligned}
\end{eqnarray}
where $\boldsymbol{1}$ represents the identity.
For simplicity, we define the following quantities
\begin{eqnarray}
\begin{aligned}
&B_x = \delta J_s\eta, B_z(k) = (-J_s+J_d)\cos (k)-\frac{\Delta}{2},\\
&l(k) = \sqrt{B^2_x+B^2_z(k)}, \alpha(k)=\frac{\delta J_s\eta}{l(k)}, \beta(k) = \frac{B_z(k)}{l(k)}.\ \ \ 
\end{aligned}
\end{eqnarray}
The elementary time evolution operator for each $k-$component reads
\begin{eqnarray}
\label{eq.time-evolution-k}
\begin{aligned}
U_0^{\pm}(k) &= e^{i\phi_0(k)}\Big[\cos \theta(k)\boldsymbol{1} \\&+i\left(\pm\alpha(k)\sigma_x+\beta(k)\sigma_z\right)\sin \theta(k)\Big], \ \ 
\end{aligned}
\end{eqnarray}
where $\phi_0(k) = T\left[(-J_s-J_d)\cos(k)+\frac{\Delta}{2}\right]$ and $\theta(k) = Tl(k).$  By defining 
\begin{eqnarray}
\label{eq.C_0}
\begin{aligned}
    C_{0,0} &= \cos \theta(k),\\ C_{0,1} &= \alpha(k) \sin \theta(k),\\ C_{0,2} &= \beta(k) \sin \theta(k),
\end{aligned}
\end{eqnarray}
we can also rewrite the operator as
\begin{eqnarray}
\label{eq.U0}
U_{0}^{\pm}(k)=\left(C_{0, 0}\boldsymbol{1} +i\left(\pm C_{0,1} \hat{\sigma}_{x}+C_{0, 2} \hat{\sigma}_{z}\right)\right)e^{i\phi_0}.
\end{eqnarray}

\paragraph{Time evolution operator.--}
Here we want to derive the multipolar operator for a nonzero integer $n$. We employ the following Ansatz for multipolar order $n$
\begin{equation}
\label{Eq.ansatz}
    \begin{aligned}
U_{n}^{\pm}=(C_{n, 0}\boldsymbol{1} +i\left({\pm}C_{n,1} \hat{\sigma}_{[n]}+C_{n, 2} \hat{\sigma}_{z}\right))e^{i\phi_n},
\end{aligned}
\end{equation}
where $[n]=x$ for even $n$, $[n]=y$ for odd $n$, and all coefficients are real and dependent on $k$. We also require $U^+_n(U_n^+)^{\dagger} = \boldsymbol{1} $ to ensure unitarity, which leads to the condition
\begin{equation}
     \left(C_{n,0}^2+C_{n,1}^2+C_{n,2}^2\right)\boldsymbol{1}+C_{n,1}C_{n,2}
     \left\{\hat{\sigma}_{[n]}, \hat{\sigma}_z\right\} = \boldsymbol{1}.
\end{equation}
As the second contribution is zero, we obtain the constraint
\begin{equation}
    C_{n,0}^2+C_{n,1}^2+C_{n,2}^2 = 1.
\end{equation}
We will use the recursive relation for multipolar operators
\begin{equation}
    U^+_{n+1} = {U}^-_nU^+_n, {U}^-_{n+1} = {U}^+_n{U}^-_n,
\end{equation}
to determine the concrete form of the coefficients $C_{n,i}$.
To start, we insert Eq.~\ref{Eq.ansatz} into the right hand side and obtain
\begin{equation}
\label{eq.expansion}
\begin{aligned}
    U^+_{n+1} &= {U}^-_n U^+_n =C_{n,{0}}^{2}\boldsymbol{1}+2 i C_{n,0} C_{n, 2} \hat{\sigma}_{z}\\
&=\left(C_{n, 0}^{2}+C_{n, 1}^{2}-C_{n, 2}^{2}\right)\boldsymbol{1}+2 i C_{n,0} C_{n,2}\hat{\sigma}_{z}\\
&+(-1) C_{n, 2} C_{n,1} 2 i \hat{\sigma}_{[n+1]} \varepsilon_{z,[n],[n+1]},
\end{aligned}
\end{equation}
where $\varepsilon_{z,[n],[n+1]}$ is the Levi-Civita function. Therefore, $\varepsilon_{z,[n],[n+1]}= 1$ for even $n$, and $\varepsilon_{z,[n],[n+1]}= -1$ for odd $n$. 
On the other hand, we also have the Ansatz
\begin{equation}
\label{Eq.ansatz_new}
    \begin{aligned}
U^+_{n+1}=C_{n+1, 0}\boldsymbol{1} +i\left(C_{n+1,1} \hat{\sigma}_{[n+1]}+C_{n+1, 2} \hat{\sigma}_{z}\right).
\end{aligned}
\end{equation}
By matching the coefficients in Eq.~\ref{Eq.ansatz_new} and Eq.~\ref{eq.expansion}, we get a set of recursive equations for coefficients
\begin{equation}
    \begin{aligned}
    \label{eq.C_n_SM}
    C_{n+1,0} & = 1-2C_{n,2}^2,\\
    C_{n+1,1} &= -2C_{n,2}C_{n,1}\varepsilon_{z,[n],[n+1]},\\
    C_{n+1,2} &= 2C_{n,0}C_{n,2}
    \end{aligned}
\end{equation}
as shown in Eq.~\ref{eq.C_n} of the main text. As such, all coefficeints in Eq.~\ref{Eq.ansatz} can be determined by using Eq.~\ref{eq.C_0} and Eq.~\ref{eq.C_n_SM}.

\paragraph{Quantum channel.--}
We are interested in the dynamics in each k-subspace after averaging over many different $n-$RMD realizations. The resultant state is not a pure state but a mixed state. We denote such a state at stroboscopic time $t=m2^nT$ as $\varrho_k^m$. Its stroboscopic time evolution is governed by
\begin{equation}
\label{eq.randomchanel}
\begin{aligned}
        \varrho^{m+1}_k = \Lambda_n(\varrho^{m}_k)&=\frac{1}{2}U_n^+\varrho^m_k (U_n^+)^\dagger+\frac{1}{2}{U}^-_n\varrho^m_k (U^-_n)^\dagger.
\end{aligned}
\end{equation}
Note, the phase $e^{i\phi_n}$ (Eq.~\ref{Eq.ansatz}) is canceled by its complex conjugate. As the structure of the quantum channel $\Lambda_n$ is independent of $k$, from now on we neglect the index $k$ for simplicity.
By inserting Eq.~\ref{Eq.ansatz} into the above expression, this channel can be rewritten as
$$\Lambda_n(\varrho)=\sum_{i,j}\mu_{ij}\hat{\sigma}_i\rho \hat{\sigma}_j,$$ with the matrix 
\begin{equation}
 \mu =  \begin{bmatrix}
    C_{n,0}^2 & 0 & -iC_{n,0}C_{n,2}  \\
    0 & C_{n,1}^2 & 0   \\
    iC_{n,0}C_{n,2} & 0 & C_{n,2}^2  \\
\end{bmatrix},
\end{equation}
with the basis $\boldsymbol{1},\hat{\sigma}_{[n]},\hat{\sigma}_z$. Note, for a fixed $n$, $\mu$ is a matrix of dimension $3$ smaller than the dimension $4$ of the full dimension for a two-level system. 
Furthermore, as $\hat{\sigma}_{[n]}$ is decoupled from other operators, we can easily diagonalize this matrix as $\mu V=VD$
\begin{equation}
\label{eq.eigenvector}
    V=\left(\begin{array}{ccc}
\frac{i C_{n,2}}{C_{n,0}} & \frac{-i C_{n,0}}{C_{n,2}} & 0 \\
0 & 0 & 1 \\
1 & 1 & 0
\end{array}\right)
\end{equation}
with the diagonal matrix
\begin{equation}
\label{eq.eigenvalue}
D =  \begin{bmatrix}
 0& 0 & 0  \\
 0& 1-C_{n,1}^2 & 0 \\
 0& 0& C_{n,1}^2\\
\end{bmatrix}.
\end{equation}
The quantum channel $\Lambda_n$ can also be expressed via Kraus operators~\cite{nielsen2002quantum}
\begin{equation}
  \Lambda_n(\varrho)= \sum_{i=1,2,3}F_i \varrho F_i^{\dagger} d_i.
\end{equation}
For $i=1$, the first eigenvalue of $\mu$ reads $d_1=0$, suggesting that $F_1$ does not contribute to the dynamics. For $i=2,3$, we have 
\begin{equation}
\begin{aligned}
      F_2 &= i\sin(\gamma)\boldsymbol{1}+\cos(\gamma)\hat{\sigma}_z, d_2 = 1-C_{n,1}^2,\\
      F_3 &= \hat{\sigma}_x, d_3 = C_{n,1}^2,
\end{aligned}
\end{equation}
where $\gamma$ is a function of $C_{n,0}$ and $C_{n,2}$. Suppose we start from the density matrix 
\begin{equation}
\label{eq.eigenvector}
    \varrho_0=\left(\begin{array}{cc}
1 & 0 \\
0 & 0
\end{array}\right),
\end{equation} one can verify that this state remains invariant under the channel $F_2$ as it is diagonal. However, $F_3$ flips the spin as
\begin{equation}
    F_3\varrho_0 F_3^{\dagger} = \left(\begin{array}{cc}
0 & 0 \\
0 & 1
\end{array}\right)=\boldsymbol{1}-\varrho_0,
\end{equation}
which physically leads to particle excitation to the d-band.
Thus, one obtains the state after time window $2^nT$ as
\begin{equation}
    \Lambda_n(\varrho_0) = (d_2-d_3)\varrho_0+d_3\boldsymbol{1}. 
\end{equation}
We can further determine the density matrix after $m$ steps by induction using the 
Ansatz\begin{equation}
\label{eq.ChannelAnsatz}
    \Lambda^m_n(\varrho_0) = f_m\varrho_0+g_m\boldsymbol{1}. 
\end{equation}
Applying one more channel to Eq.~\ref{eq.ChannelAnsatz} leads to
\begin{equation}
\begin{aligned}
\label{eq.ChannelEquation}
        \Lambda^{m+1}_n(\varrho_0) &= f_{m}\Lambda_n(\varrho_0)+g_{m}\Lambda_n(\boldsymbol{1})\\
        &=f_m(d_2-d_3)\varrho_0 + (d_3f_m+g_m)\boldsymbol{1}.
\end{aligned}
\end{equation}
By matching the coefficients in Eq.~\ref{eq.ChannelAnsatz} and Eq.~\ref{eq.ChannelEquation}, we have the recursive equation
\begin{equation}
    f_{m+1} = f_m(d_2-d_3),\ g_{m+1} = d_3f_m+g_m,
\end{equation}
which can be solved for arbitrary $m$
\begin{equation}
    f_m = (d_2-d_3)^m,\ g_m = d_3 \frac{1-(d_2-d_3)^m}{1-(d_2-d_3)}.
\end{equation}
Therefore, after $m$ steps, the density matrix at stroboscopic time $t=m2^nT$ reads
\begin{eqnarray}
\begin{aligned}
        \Lambda^m(\varrho_0) &= \left(\begin{array}{cc}
(d_2-d_3)^m+d_3\frac{1-(d_2-d_3)^m}{1-(d_2-d_3)} & 0 \\
0 & d_3\frac{1-(d_2-d_3)^m}{1-(d_2-d_3)}
\end{array}\right)\\&=\left(\begin{array}{cc}
\frac{1+(1-2C^2_{n,1})^m}{2} & 0 \\
0 & \frac{1-(1-2C^2_{n,1})^m}{2}.
\end{array}\right)
\end{aligned}
\end{eqnarray}
which represents a classical mix of spin up or down. After sufficiently long time, because $0<1-2C_{n,1}^2<1$,  $(1-2C_{n,1}^2)^{\infty}\to 0$ for  $m\to\infty$. Eventually, we have 
\begin{equation}
    \Lambda^m(\varrho_0) \to \left(\begin{array}{cc}
0.5 & 0 \\
0 & 0.5
\end{array}\right),
\end{equation} suggesting that the originally polarized state now has a equal distribution between spin up or down

\paragraph{Incoherent sum over $k-$components.--}
For a single particle Fock state at site $l=0$ in the $s-$band $
\ket{\psi(0)} = \hat{b}_{s0}^{\dagger}\ket{0} = \frac{1}{\sqrt{L}}\sum_k\hat{b}^{\dagger}_{sk}\ket{0},
$ its density matrix contains both the diagonal 
\begin{eqnarray}
\varrho_k^0 = \frac{1}{L} \hat{b}_{sk}^{\dagger}\ket{0}\bra{0}\hat{b}_{sk}
\end{eqnarray}
and off-diagonal contributions between different quasi-momentum components. These off-diagonal contributions tend to average out for the RMD protocol, hence, the time averaged dynamics can be well approximated by the incoherent sum over all $k$. 

For a single momentum $k$, the particle loss time $\tau_{\mathrm{loss}}$ diverges if the $k-$component completes a full Rabi cycle at stroboscopic times when $C_{n,1}(k)=0$. For instance, when $n=0$, according to Eq.~\ref{eq.C_0}, we need $\sin\theta(k)=0$ with $\theta(k)=Tl(k)$. It leads to the condition $Tl(k)=q\pi$ for an integer $q$, hence, the $k-$component can complete a Rabi cycle when the driving period $T$ matches the Rabi oscillation periods
\begin{eqnarray}
\label{eq.Tk}
T_k = \frac{q\pi}{l(k)},
\end{eqnarray}
with $l(k)=\sqrt{(\delta J_s\eta)^2+[(-J_s+J_d)\cos(k)-\Delta/2]^2}$. If $J_s=J_d$, $l(k)$ becomes a constant, but for more general cases, it depends on $k$. Consequently, the driving period $T$ cannot match $T_k$ exactly for all $k-$components at the same time. 

However, deviations between different $T_k$ can be small if the difference between the intra-band hopping rates is much smaller than the band gap $(J_d-J_s)\ll \Delta$, so that $T$ can match $T_k$ approximately. To see this, one can Taylor expand Eq.~\ref{eq.Tk} as
\begin{eqnarray}
\begin{aligned}
{T_k} = \frac{q\pi}{\sqrt{(\Delta/2)^2+(\delta J_s\eta)^2}}\Big[1+\frac{2\cos (k)(J_d-J_s)}{\Delta}\\+\mathcal{O}\Big((\frac{J_d-J_s}{\Delta})^2\Big)\Big].\ \ 
\end{aligned}
\end{eqnarray}
The leading order contribution to the deviation of these periods scales as $ (J_d-J_s)/\Delta$. 
As such, for \begin{eqnarray}
\label{eq.condition}
(J_d-J_s)\ll\Delta,
\end{eqnarray}  starting from the state $\ket{\psi(0)}$, the particle can still approximately complete a Rabi-like cycle between the two bands if $T\approx T_k$. Once it happens, the inter-band heating can be significantly suppressed, although now the particle loss time $\tau_{\mathrm{loss}}$ is always finite. For a deep lattice, $J_s$ can be smaller than $\Delta$ by two orders of magnitude, hence, Eq.~\ref{eq.condition} simply reduces to $J_d\ll\Delta$.

One can therefore use this idea to suppress the particle loss at stroboscopic times.
We compute the particle number in the s-band
\begin{eqnarray}
\label{eq.NSdynamics}
\begin{aligned}
N_{\mathrm{s}}(2^nTm) = \sum_k \frac{1+(1-2C_{n,1}^2(k))^m}{2}\frac{1}{L},
\end{aligned}
\end{eqnarray}
as given in the main content.  By requiring $N_{\mathrm{s}}(\tau_{\mathrm{loss}})/N_{\mathrm{tot}}=N_c$, one can extract the particle loss time $\tau_{\mathrm{loss}}$. For different values of $m$, an individual sum over all $k-$components is needed. It can be simplified if $C_{n,1}^2$ is small, which happens for $\delta J_s\eta\ll\Delta$. One can Taylor expand Eq.~\ref{eq.NSdynamics} to obtain the simplified expression
\begin{eqnarray}
\begin{aligned}
\label{eq.NsSimple}
N_{\mathrm{s}}(2^nTm) &\approx \sum_k \frac{2-2mC_{n,1}^2(k)}{2}\frac{1}{L}\\
&=1-m\sum_k \frac{C_{n,1}^2(k)}{L},
\end{aligned}
\end{eqnarray}
for which only a single sum over different $k-$components is required. This approximation works well when $N_{\mathrm{s}}$ remains close to 1, and in Fig.2 (b), we use $N_c=0.92$ to extract the particle loss time $\tau_{\mathrm{loss}}$.

As shown in Fig.2 (b), for instance around $1/T\sim35$, $\tau_{\mathrm{loss}}$ exhibits a well-pronounced peak suggesting that particle loss is remarkably suppressed. However, instead of diverging, its maximum value now becomes finite.

  \begin{figure}
	\includegraphics[width=0.99\linewidth]{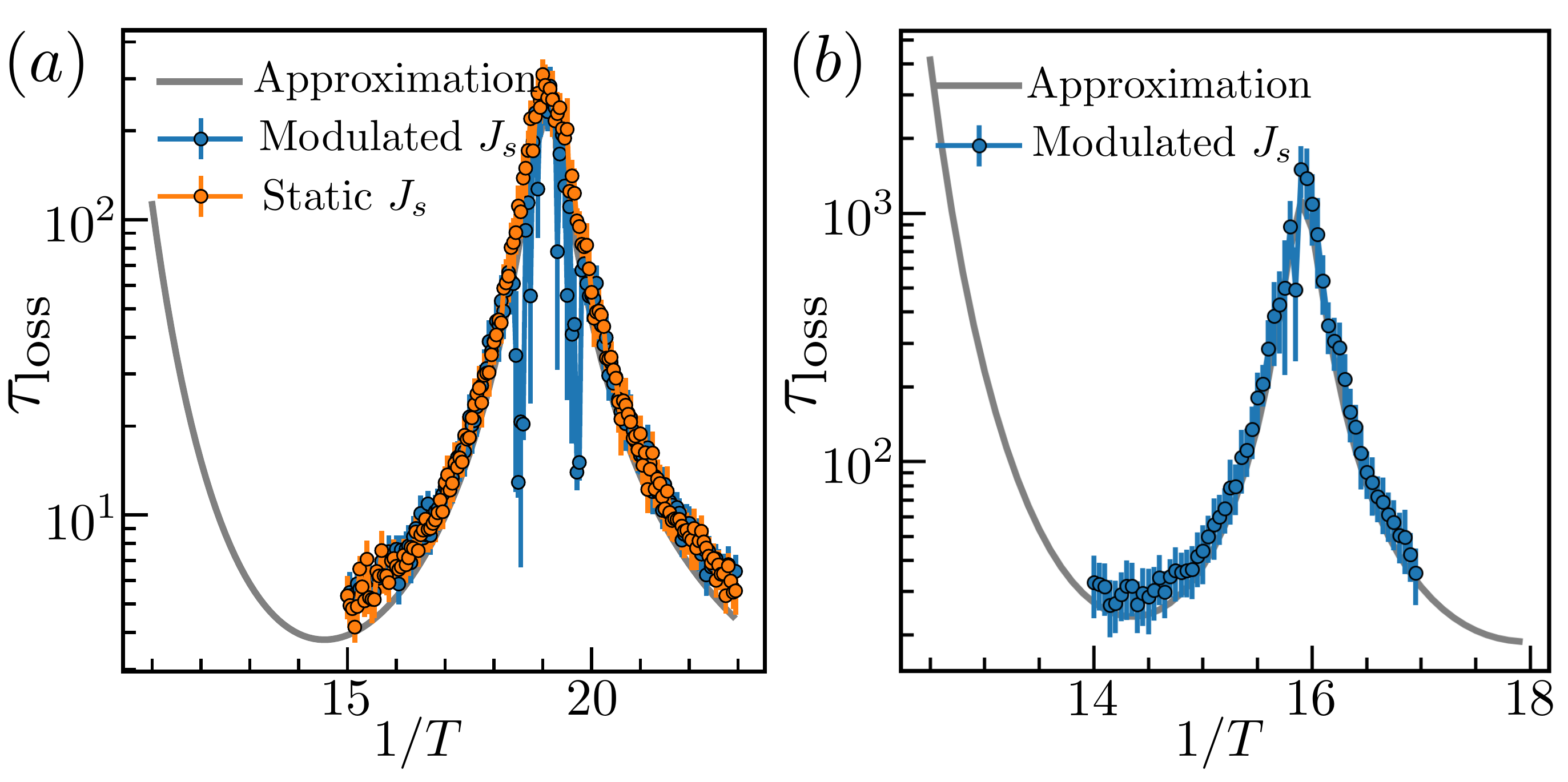}
	\caption{ Dependence of the particle loss time scale $\tau_{\mathrm{loss}}$ on the driving frequency $1/T$ for the $2-$RMD for band gap $\Delta=60$ and $\Delta=150$ in panel (a) and (b) respectively. We use $\delta J=2,J_d=2,\mu=10, L=20,N_{\mathrm{tot}}=1$ for numerical simulation.}
	\label{fig:scaling_hopping}
\end{figure}
\section{Dependence on the band gap}
As shown above, to make the system analytically tractable we neglected all interactions, the modulation in s-band, as well as the staggered potential. This approximation works well when the band gap is much larger than all other energy scales. However, for a smaller band gap, those neglected processes may still be visible.  As shown in Fig.~\ref{fig:scaling_hopping} (a), by explicitly excluding the modulation in the hopping rate  $J_s$ of the s-band, the numerical results (orange) can be accurately captured by our analytical theory. However, once we use a large modulation in $J_s$ (blue), although the analytical solution still works nicely for most frequencies, notable deviations can be observed in the regime $18<1/T<20$. These deviations become negligible if we increase the band gap as shown in the panel (b).

\section{Scaling for dipolar $1-$RMD}
\begin{figure}
	\includegraphics[width=0.99\linewidth]{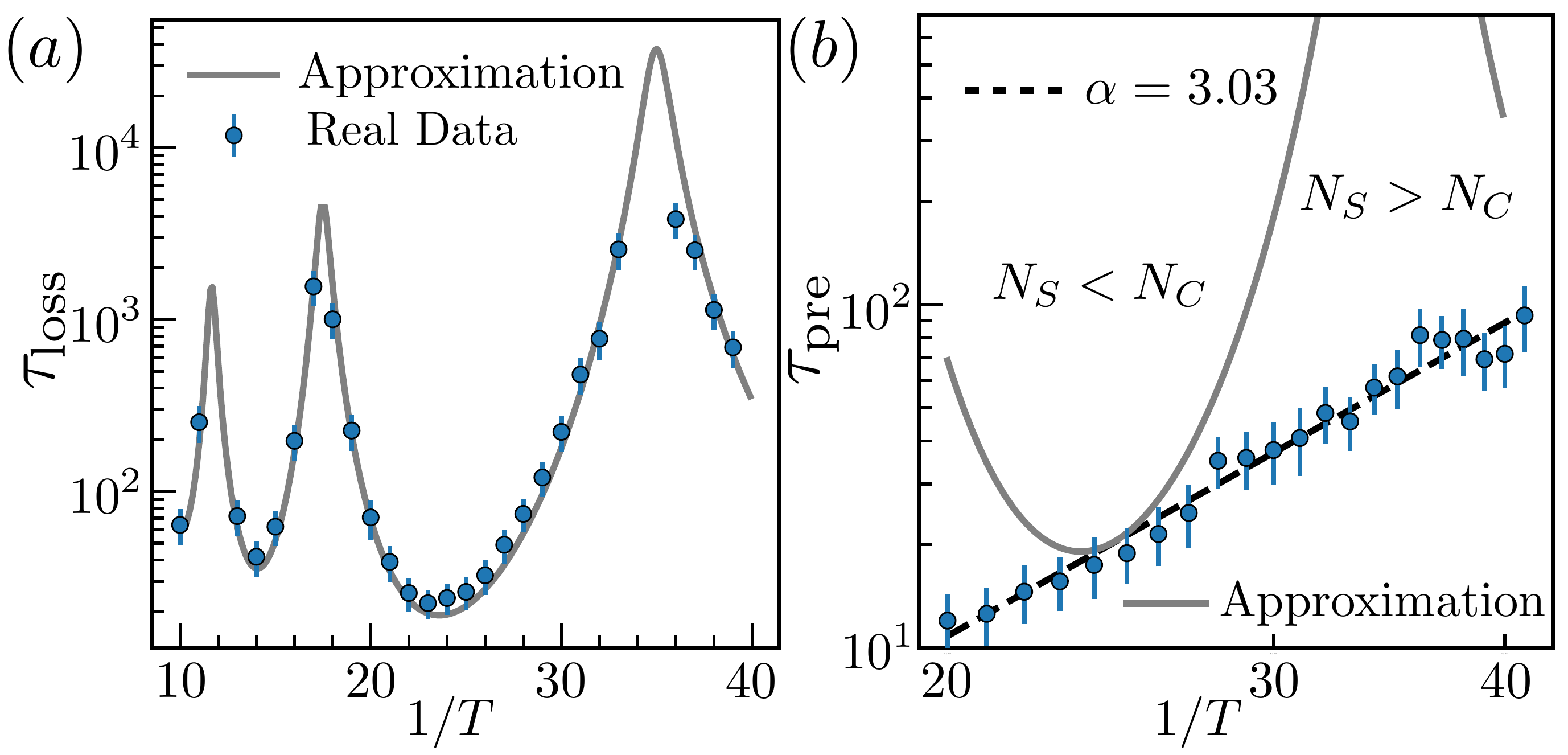}
	\caption{(a) Dependence of the particle loss time scale $\tau_{\mathrm{loss}}$ on the driving frequency $1/T$ for the $1-$RMD. Similar to the $2-$RMD in the main content, there exist several frequency windows where excitations to the d-band are significantly suppressed.  (b) In general, $\tau_{\mathrm{pre}}$ is shorter than $\tau_{\mathrm{loss}}$ for $n=1$, therefore, the prethermal lifetime scales algebraically with exponent $3$ for all frequencies. We use $\delta J=1.2,J_d=7,U_s=5,U_d=3,U_{sd}=2,\mu=10,\Delta=220$ for numerical simulation.}
	\label{fig:scaling_n1}
\end{figure}

In Fig.~\ref{fig:scaling_n1}, we show the dependence of the two heating times for the diplar $1-$RMD.
The particle loss time scale $\tau_{\mathrm{loss}}$ is determined by averaging the time when $N_{\mathrm{s}}(t)/N_{\mathrm{tot}}$ drops below $0.9,0.9\pm0.03,0.9\pm0.015$. The prethermal lifetime
$\tau_{\mathrm{pre}}$ is determined by $I(t)=0.55,0.55\pm0.08,0.55\pm0.04$. Particle loss is negligible and the scaling exponent of the prethermal lifetime fits well with the predicted value $2n+1$.

\end{document}